\documentclass[
twocolumn,
]{article}

\usepackage[utf8]{inputenc} 

\usepackage[T1]{fontenc} 

\usepackage{graphicx} 

\usepackage[
  backend=biber,
  url=true,
  style=numeric,
  sorting=none, 
  maxnames=2,
  minnames=1,
  maxbibnames=99,
  giveninits,
  uniquename=init
  ]{biblatex} 

\usepackage[braket, qm]{qcircuit} 

\usepackage{chemformula} 
\setchemformula{
    format = \rmfamily,
}

\usepackage[hidelinks]{hyperref} 

\usepackage{cleveref}
\crefformat{subsection}{Subsection~#2#1#3}
\Crefformat{subsection}{Subsection~#2#1#3}

\usepackage{orcidlink} 
\usepackage{tabto} 

\usepackage{siunitx} 
\usepackage{bbold} 
\usepackage{mathtools} 
\usepackage{amssymb}
\usepackage{tikz}
\usepackage{cuted} 
\usepackage{bbm} 
\usepackage{enumitem} 

\usepackage{subcaption}
\usepackage{caption}
\usepackage{booktabs}
\usepackage{stfloats}
\usepackage{placeins}
\usepackage{float}
\usepackage{multirow} 
\usepackage[shortcuts,acronym]{glossaries} 
\setkeys{glslink}{hyper=false} 

\newacronym{ARP}{ARP}{Adiabatic Rapid Passage}
\newacronym{CB}{CB}{Conduction Band}
\newacronym{CPTP}{CPTP}{completely positive, trace-preserving}
\newacronym{CW}{CW}{Continuous Wave}
\newacronym{DPE}{DPE}{Dichromatic Pulsed Excitation}
\newacronym{FSS}{FSS}{Fine Structure Splitting}
\newacronym{GKSL}{GKSL}{Gorini-Kossakowski-Sudarshan-Lindblad}
\newacronym{LA}{LA}{Longitudinal-Acoustic} 
\newacronym{MAC}{MAC}{Message Authentication Code}
\newacronym{QC}{QC}{Quantum Channel}
\newacronym{QD}{QD}{Quantum Dot}
\newacronym{QKD}{QKD}{Quantum Key Distribution}
\newacronym{TPE}{TPE}{Two-Photon Excitation}
\newacronym{VB}{VB}{Valence Band}

\newcommand{\ie}{\textrm{i.e.}~} 

\newcommand{\Ie}{\textrm{I.e.}~} 

\newcommand{\eg}{\textrm{e.g.}~} 

\newcommand{\viz}{\textrm{viz.}~} 

\DefineBibliographyStrings{english}{%
    andothers = {\textrm{et al\adddot}}
}

\newcommand{\HQD}{\ensuremath{\mathcal{H}_{\text{QD}}}}
\newcommand{\ketG}{\ensuremath{\ket{\mathrm{G}}}}

\newcommand{\ketXtwo}{\ensuremath{\ket{\mathrm{X_{2}}}}}
\newcommand{\ketXone}{\ensuremath{\ket{\mathrm{X_{1}}}}}
\newcommand{\ketXX}{\ensuremath{\ket{\mathrm{XX}}}}
\newcommand{\sigmaOp}[2]{\ensuremath{\hat{\sigma}_{#1 \to #2}}}

\bibliography{bibliography}

\hypersetup{
  pdftitle={Entangled Photon Pair Generator via Biexciton-Exciton Cascade in Semiconductor Quantum Dots and its Simulation},
  pdfauthor={Simon Sekavčnik and Paul Kohl and Janis Nötzel},
  pdfsubject={Biexciton-Exciton Cascade in Semiconductor Quantum Dots},
  pdfkeywords={Entangled Photon Generation, Semiconductor Quantum Dots, Biexciton-Exciton Cascade}
}

\newcommand{\Mark}[1]{\textsuperscript{#1}}

\begin{document}

\twocolumn[{ 
    \centering
    \LARGE Entangled Photon Pair Generator via Biexciton-Exciton Cascade in Semiconductor Quantum Dots and its Simulation \\[1cm]
    \large Simon Sekavčnik\Mark{$\dagger$, *} \orcidlink{0000-0002-1370-9751},
           Paul Kohl\Mark{$\dagger$, *} \orcidlink{0009-0000-6990-0515},
           Janis Nötzel\\[0.5cm]
    \normalsize
    \Mark{$\dagger$} These authors contributed equally to this work.\\[0.25cm]
    \Mark{*} Corresponding authors. E-Mail: \url{simon.sekavcnik@tum.de}, \url{paul.kohl@tum.de}.\\[0.25cm]
    
    \begin{tabular}{*{1}{c}}
        S. Sekavčnik, P. Kohl, J. Nötzel \\
        Emmy-Noether Group Theoretical Quantum Systems Design,\\
        TUM School of Computation, Information and Technology\\ 
        Technical University of Munich (TUM), \\ 
        Theresienstraße 90, München, 80333, Bavaria, Germany. \\ 
    \end{tabular}\\[1cm]

    \\[1cm]
    
}]


\begin{abstract}

The generation of entangled photon pairs is highly useful for many types of quantum technologies. In this work an entangled photon pair generator that utilises the biexciton-exciton cascade in semiconductor quantum dots is described on a physical, mathematical, and software level. The system is implemented and simulated as a self-contained component in a framework for bigger quantum optical experiments. Thus, it is a description to further the holistic understanding of the system for interdisciplinary audiences in a hopefully simple yet sufficient manner. It is described from the condensed matter physics fundamentals, over the most important quantum optical properties, to a mathematical description of the used model, and finally a software description and simulation, making it an executable description of such a system.
We provide a compact description in the Kraus operator formalism to seamlessly incorporate such an entangled photon pair generator simulation component into bigger simulations consisting of multiple components at a reasonable computational cost. The simulation accommodates a wide range of parameter regimes and makes it possible to simulate many different excitation strategies. This includes resonant two-photon excitation, adiabatic rapid passage chirped excitation, and dichromatic pulsed excitation.

\end{abstract}


\section{Introduction}\label{section:Introduction}

The generation of entangled photon pairs is of high interest for several applications in quantum computing and quantum communication. These include \gls{QKD} based on entangled qubits \cite{EkertQKD}, \glspl{MAC} \cite{CurtyMAC} pre-sharing maximally entangled qubits to authenticate a classical bit, combinations of \gls{QKD} and Quantum Authentication \cite{ShiAuth}, and many more.

Entanglement may enhance the communication performance as well. Entanglement-assisted communication \cite{sekavcnik2023scaling} is an active research direction investigating the speedup that pre-shared entanglement can bring to telecommunications.

Yet, there seems to be an inherent disconnect between theory, experiment, and application in the sense that the different fields may use different modelling techniques, focus on different aspects, and work with different mathematical formalisms.
This makes the integration of different components in a realistic way often quite challenging. The fundamental research on physical devices is done by specialised experimental research groups, channel capacities and channel models are developed by theorists under (more or less realistic) assumptions, and engineering may not only be tasked with the design of technical application of devices and ensuring their usability and robustness, but also scaling them to commercial products. These different tasks inescapably yield difficulties in the transfer of knowledge at the intersections.

The main focus of this work is an attempt to remedy this. In our view a big library of components that are realistic, yet not too specialised is necessary to remotely achieve this goal. To that end we begin with a component that is central to many (photonic) quantum technologies: An entangled photon pair generator.

Thus, in this work we describe an entangled photon generator in the form of a biexciton-exciton cascade that can be found in semiconductor \glspl{QD} and provide a mathematical description and a software simulation of it. The simulation consists of a set of self-contained source files in Python that can be run on commodity hardware for preliminary design of quantum experiments. They are easily extensible and can be readily integrated with other such components.

This work is structured as follows: Section \ref{section:Physics} introduces preliminary notions from condensed matter physics and describes the physical structure of a biexciton-exciton cascade based on \glspl{QD} with some commonly used excitation schemes.
Section \ref{section:Mathematics} dwells on the mathematical understanding of such a system and provides a mathematical model of the latter. The quantum channel formalism is employed to develop a compact component from this model as Kraus operators. This component is meant to be used in combination with multiple other components for other physical devices.
This approach enables the user of the simulation to connect different realistic simulation components developed by specialised research groups in one framework with minimal effort.
Section \ref{section:Simulation} provides details on our classical simulation of the biexciton-exciton cascade and discussion of its results.
Finally, in Section \ref{section:Conclusion} we draw our conclusions and outline possible future developments.

\section{Physics}\label{section:Physics}

In order to understand how the biexciton-exciton cascade energy structure -- which may be used for entangled photon pair generation -- is achieved, it is instructive to recapitulate the relevant concepts from condensed matter physics.

\subsection{Bandgap Structure}

A large enough periodic lattice of atoms or molecules creates an electronic band structure for the electrons in the lattice from the collection of quantum mechanically allowed and disallowed energy states at each point of the lattice. 

The band structure emerges as follows: 
In the quasi-infinite periodic lattice of atoms or molecules of the bulk material each of the electrons would individually exhibit discrete energy states allowed to occupy. But in the lattice not only the energy levels determined by the potential of one single particle under consideration can be occupied, because also the neighbouring particles (in all directions) have a non-negligible influence on the potential at the position under consideration.

Due to the periodic nature of the bulk crystal structure also the energy levels have a periodic structure w.r.t.~the position in the crystal. If the lattice is quasi-infinite \footnote{\Ie large enough that the particle furthest from the considered position has only negligible influence.}, the discrete energy levels of allowed states are so close together that they form approximately continuous bands of allowed states (energy bands) and disallowed energy states (energy gaps/bandgaps). When including the electrons occupying their respective levels, this enables us to determine the crystal's macroscopic behaviour and sort materials into the categories of metal, semimetal, p-type semiconductor, intrinsic semiconductor, n-type semiconductor, and insulator, depending on where the \textit{Fermi level}\footnote{The Fermi level is the (hypothetical) energy level at which the probability that it is occupied by electrons is $\frac{1}{2}$. \cite[p.136, p.205]{IntroSolidStatePhysics}} is located. \cite[pp.161 et seqq.]{IntroSolidStatePhysics}

The \gls{VB} is the band located below the Fermi level while the \gls{CB} is the band above the Fermi level. The energy bands of the crystal up to the \gls{VB} are filled with "bound" electrons, while the \gls{CB} contains the "free" electrons, thus the conductive behaviour of materials can be described with the \gls{VB} and \gls{CB}. In order to distinguish different materials one looks at the occupancy of the energy bands with electrons at a certain thermodynamic equilibrium state under consideration. 

For a \textit{p-type semiconductor} the Fermi Level is located \textit{in between two energy bands, but closer to the \gls{VB} and the bands are relatively close together}, that means that the \gls{VB} is fully occupied with a moderate probability and the \gls{CB} is empty with very high probability.
For an \textit{intrinsic semiconductor} the Fermi Level is located \textit{roughly equidistantly in between two energy bands and the bands are relatively close together}, that means that the \gls{VB} is fully occupied with high probability and the \gls{CB} is empty with high probability.
For an \textit{n-type semiconductor} the Fermi Level is located \textit{in between two energy bands, but closer to the \gls{CB} and the bands are relatively close together}, that means that the \gls{VB} is fully occupied with a very high probability and the \gls{CB} is empty with moderate probability. \cite{IntroSolidStatePhysics}

Thus, there are following charge carriers in a solid: the electron $\mathrm{e}^-$ with a negative charge and a specific spin (which is free as in the \gls{CB} or bound), and if in the filled "electron sea" in the \gls{VB} an electron is missing, this behaves itself like a particle which carries the opposite charge of an electron and a spin in the opposite direction to the spin of the missing electron. This is called the hole $\mathrm{h}^+$.

\subsection{Semiconductor Quantum Dots}

If one can now confine charge carriers -- or more specifically their wave function -- in a potential well-like structure one gets discrete energy levels again, giving rise to quantum mechanical behaviour. This is done \eg in a semiconductor \gls{QD} by encasing a semiconductor material with a relatively small bandgap in another semiconductor material with a higher bandgap to create a potential barrier on the material interfaces. This confines the wavefunction of the charge carriers in all three spatial dimensions, yielding a point-like quantum structure, which is reflected in the term \textit{quantum dot}.
\cite{ThesisOpticalCharacterisationOfTelecommunication-WavelengthQDs}

These discrete levels are occupied as follows in the ground state: The discrete levels inside the \gls{VB} are completely filled with $\mathrm{e}^-$, \ie they are devoid of $\mathrm{h}^+$; The discrete levels within the \gls{CB} are completely devoid of $\mathrm{e}^-$. Summarised, this is called "an empty \gls{QD}" in the following.
If now one would excite one $\mathrm{e}^-$ from the \gls{VB} into the \gls{CB}, this is the same as creating a $\mathrm{h}^+$ in the \gls{VB} and creating an $\mathrm{e}^-$ in the \gls{CB}, which are coupled and can recombine. This leads to the concept of excitonic quasi-particles, which will be discussed now.

\subsection{Excitonic Quasi-Particles}\label{subsec:ExcitonicQuasi-Particles}

So called \textit{excitons} are quasi-particles consisting of pairs of an electron and a hole bound together by their Coulomb interaction. They behave like an approximation of a two-level atom. 
More specifically, the exciton $\mathrm{X}$ can be created in case the temperature is so low, that $k_\text{B} T$ is less than the electron-hole binding energy and the electron and hole can recombine radiatively, that is via emitting a photon.
The energy of the released photon is $E_\text{g} - E_{\text{b},\mathrm{X}}$, where $E_{\text{b},\mathrm{X}}$ is the exciton binding energy and $E_\text{g}$ the bandgap energy. 
\cite[p.~61]{QuantumOptics}\cite[p.~15]{DissZeuner}

There are different kinds of excitonic quasi-particles, their properties are dependent on how many $\mathrm{e}^-$ are excited in the \gls{CB} and how many $\mathrm{h}^+$ are located in the \gls{VB} bound to the respective electrons, the spin of both electrons and holes involved, and their effective masses\footnote{The effective mass describes how particles react to forces if they had this mass.}.
A \gls{QD} exhibits an electronic structure analogous to the s-shell of an atom. The s-shell has a degeneracy of $2$, which means it can be occupied by two $\mathrm{e}^-$ of opposite spin according to the \textit{Pauli Principle}. This gives rise to different possible occupancies shown in Figure \ref{fig:ExcitonicQuasi-Particles}, where we just take into account heavy holes (\ie those with more effective mass), because light holes are weakly bound due to their smaller effective mass and compressive tension in \glspl{QD}. 

These combinations form the following quasi-particles: Neutral excitons -- commonly just called "excitons" -- come in two flavours, \viz so-called (bright) excitons consisting of one $\mathrm{e}^-$ and one $\mathrm{h}^+$ of opposite spin, denoted $\mathrm{X}$, and so-called dark excitons denoted $\mathrm{DX}$ where $\mathrm{e}^-$ and $\mathrm{h}^+$ have the same spin direction. $\mathrm{DX}$ are optically not active and $\mathrm{X}$ are optically active. This is because the spin of $\mathrm{e}^-$ is $\pm\frac{1}{2}$ and the projection of total angular momentum of heavy holes is $\pm\frac{3}{2}$ and thus $\mathrm{X}$ has total spin $\pm 1$ and circularly polarised light also has spin $\pm 1$. The $\mathrm{DX}$ on the other hand has spin $\pm 2$. Thus conservation of angular momentum dictates that light couples to $\mathrm{X}$ and not $\mathrm{DX}$. Additionally, there are the negative Trions $\mathrm{X}^-$ consisting of two excited $\mathrm{e}^-$ and one $\mathrm{h}^+$, and the positive Trions $\mathrm{X}^+$ consisting of one excited $\mathrm{e}^-$ and two $\mathrm{h}^+$. If there are two excited $\mathrm{e}^-$ and also two $\mathrm{h}^+$ then this constitutes the so called Biexciton $\mathrm{XX}$.
The reasoning regarding conservation of angular momentum also applies to $\mathrm{X}^+$, $\mathrm{X}^-$, and $\mathrm{XX}$, which has the consequence that they are also optically active.
\cite{DissSimonGordon, DissZeuner, ThesisOpticalCharacterisationOfTelecommunication-WavelengthQDs}

\begin{figure}[!htbp]
    \centering
    \begin{tikzpicture}[
            scale = 0.5,
            every node/.style={scale=1.00},
            OuterBox/.style={gray, rectangle, densely dashed, minimum width=1.6cm, minimum height=1.6cm, draw},
            OuterBoxSmall/.style={gray, rectangle, densely dashed, minimum width=0.8cm, minimum height=1.6cm, draw},
            VB/.style={black, rectangle, minimum width=1.6cm, minimum height=0.8cm},
            VBSmall/.style={black, rectangle, minimum width=0.8cm, minimum height=0.8cm},
            CB/.style={black, rectangle, minimum width=1.6cm, minimum height=0.8cm},
            CBSmall/.style={black, rectangle, minimum width=0.8cm, minimum height=0.8cm},
        ]
        
        \node[OuterBox, name=a] at (-6,0) {};
        \node[OuterBox, name=b] at (-2.6,0) {};
        \node[OuterBox, name=c] at (0.8,0) {};
        \node[OuterBox, name=d] at (4.2,0) {};
        
        \node[OuterBoxSmall, name=e] at (6.8,0) {};

        \draw (a.south) node[below] {a)};
        \draw (b.south) node[below] {b)};
        \draw (c.south) node[below] {c)};
        \draw (d.south) node[below] {d)};
        \draw (e.south) node[below] {e)};

        \draw[dash dot, gray] (a.south) -- (a.north);
        \draw[dash dot, gray] (b.south) -- (b.north);
        \draw[dash dot, gray] (c.south) -- (c.north);
        \draw[dash dot, gray] (d.south) -- (d.north);

        \node[VB, name=aVB] at (a.south) {}; 
        \node[VB, name=bVB] at (b.south) {}; 
        \node[VB, name=cVB] at (c.south) {}; 
        \node[VB, name=dVB] at (d.south) {}; 
        
        \node[VBSmall, name=eVB] at (e.south) {};

        \draw (aVB.148) -- (aVB.110);
        \draw (aVB.70)  -- (aVB.32);
        \draw (bVB.148) -- (bVB.110);
        \draw (bVB.70)  -- (bVB.32);
        \draw (cVB.148) -- (cVB.110);
        \draw (cVB.70)  -- (cVB.32);
        \draw (dVB.148) -- (dVB.110);
        \draw (dVB.70)  -- (dVB.32);
        
        \draw (eVB.58) -- (eVB.122);

        \node[CB, name=aCB] at (a.north) {}; 
        \node[CB, name=bCB] at (b.north) {}; 
        \node[CB, name=cCB] at (c.north) {}; 
        \node[CB, name=dCB] at (d.north) {};
        
        \node[CBSmall, name=eCB] at (e.north) {};

        \draw (aCB.-148) -- (aCB.-110);
        \draw (aCB.-70)  -- (aCB.-32);
        \draw (bCB.-148) -- (bCB.-110);
        \draw (bCB.-70)  -- (bCB.-32);
        \draw (cCB.-148) -- (cCB.-110);
        \draw (cCB.-70)  -- (cCB.-32);
        \draw (dCB.-148) -- (dCB.-110);
        \draw (dCB.-70)  -- (dCB.-32);

        \draw (eCB.-58) -- (eCB.-122);

        \node[circle, minimum size=0.5cm,   name=aVB1] at       (aVB.141) {}; 
        \node[circle, minimum size=0.5cm,   name=aVB2] at       (aVB.125) {}; 
        \draw[<-, >=stealth]                    (aVB1.south) -- (aVB1.north);
        \filldraw[fill=white, draw=black]       (aVB1) circle (0.1cm) {}; 

        \node[circle, minimum size=0.5cm,   name=aCB1] at       (aCB.-141) {}; 
        \node[circle, minimum size=0.5cm,   name=aCB2] at       (aCB.-125) {}; 
        \draw[->, >=stealth]                    (aCB1.south) -- (aCB1.north);
        \filldraw[fill=black, draw=black]       (aCB1) circle (0.1cm) {}; 

        \node[circle, minimum size=0.5cm,   name=aVB3] at       (aVB.55) {}; 
        \node[circle, minimum size=0.5cm,   name=aVB4] at       (aVB.39) {}; 
        \draw[<-, >=stealth]                    (aVB4.north) -- (aVB4.south);
        \filldraw[fill=white, draw=black]       (aVB4) circle (0.1cm) {}; 

        \node[circle, minimum size=0.5cm,   name=aCB3] at       (aCB.-55) {}; 
        \node[circle, minimum size=0.5cm,   name=aCB4] at       (aCB.-39) {}; 
        \draw[->, >=stealth]                    (aCB4.north) -- (aCB4.south);
        \filldraw[fill=black, draw=black]       (aCB4) circle (0.1cm) {};

        \node[circle, minimum size=0.5cm,   name=bVB1] at       (bVB.141) {}; 
        \node[circle, minimum size=0.5cm,   name=bVB2] at       (bVB.125) {}; 
        \draw[<-, >=stealth]                    (bVB2.north) -- (bVB2.south);
        \filldraw[fill=white, draw=black]       (bVB2) circle (0.1cm) {}; 

        \node[circle, minimum size=0.5cm,   name=bCB1] at       (bCB.-141) {}; 
        \node[circle, minimum size=0.5cm,   name=bCB2] at       (bCB.-125) {}; 
        \draw[->, >=stealth]                    (bCB1.south) -- (bCB1.north);
        \filldraw[fill=black, draw=black]       (bCB1) circle (0.1cm) {}; 

        \node[circle, minimum size=0.5cm,   name=bVB3] at       (bVB.55) {}; 
        \node[circle, minimum size=0.5cm,   name=bVB4] at       (bVB.39) {}; 
        \draw[<-, >=stealth]                    (bVB3.south) -- (bVB3.north);
        \filldraw[fill=white, draw=black]       (bVB3) circle (0.1cm) {}; 

        \node[circle, minimum size=0.5cm,   name=bCB3] at       (bCB.-55) {}; 
        \node[circle, minimum size=0.5cm,   name=bCB4] at       (bCB.-39) {}; 
        \draw[->, >=stealth]                    (bCB4.north) -- (bCB4.south);
        \filldraw[fill=black, draw=black]       (bCB4) circle (0.1cm) {};

        \node[circle, minimum size=0.5cm,   name=cVB1] at       (cVB.141) {}; 
        \node[circle, minimum size=0.5cm,   name=cVB2] at       (cVB.125) {}; 
        \draw[<-, >=stealth]                    (cVB1.south) -- (cVB1.north);
        \draw[<-, >=stealth]                    (cVB2.north) -- (cVB2.south);
        \filldraw[fill=white, draw=black]       (cVB1) circle (0.1cm) {}; 
        \filldraw[fill=white, draw=black]       (cVB2) circle (0.1cm) {}; 

        \node[circle, minimum size=0.5cm,   name=cCB1] at       (cCB.-141) {}; 
        \node[circle, minimum size=0.5cm,   name=cCB2] at       (cCB.-125) {}; 
        \draw[->, >=stealth]                    (cCB1.south) -- (cCB1.north);
        \filldraw[fill=black, draw=black]       (cCB1) circle (0.1cm) {}; 

        \node[circle, minimum size=0.5cm,   name=cVB3] at       (cVB.55) {}; 
        \node[circle, minimum size=0.5cm,   name=cVB4] at       (cVB.39) {}; 
        \draw[<-, >=stealth]                    (cVB3.south) -- (cVB3.north);
        \draw[<-, >=stealth]                    (cVB4.north) -- (cVB4.south);
        \filldraw[fill=white, draw=black]       (cVB3) circle (0.1cm) {}; 
        \filldraw[fill=white, draw=black]       (cVB4) circle (0.1cm) {}; 

        \node[circle, minimum size=0.5cm,   name=cCB3] at       (cCB.-55) {}; 
        \node[circle, minimum size=0.5cm,   name=cCB4] at       (cCB.-39) {}; 
        \draw[->, >=stealth]                    (cCB4.north) -- (cCB4.south);
        \filldraw[fill=black, draw=black]       (cCB4) circle (0.1cm) {};

        \node[circle, minimum size=0.5cm,   name=dVB1] at       (dVB.141) {}; 
        \node[circle, minimum size=0.5cm,   name=dVB2] at       (dVB.125) {}; 
        \draw[<-, >=stealth]                    (dVB1.south) -- (dVB1.north);
        \filldraw[fill=white, draw=black]       (dVB1) circle (0.1cm) {}; 

        \node[circle, minimum size=0.5cm,   name=dCB1] at       (dCB.-141) {}; 
        \node[circle, minimum size=0.5cm,   name=dCB2] at       (dCB.-125) {}; 
        \draw[->, >=stealth]                    (dCB1.south) -- (dCB1.north);
        \draw[->, >=stealth]                    (dCB2.north) -- (dCB2.south);
        \filldraw[fill=black, draw=black]       (dCB1) circle (0.1cm) {}; 
        \filldraw[fill=black, draw=black]       (dCB2) circle (0.1cm) {}; 

        \node[circle, minimum size=0.5cm,   name=dVB3] at       (dVB.55) {}; 
        \node[circle, minimum size=0.5cm,   name=dVB4] at       (dVB.39) {}; 
        \draw[<-, >=stealth]                    (dVB4.north) -- (dVB4.south);
        \filldraw[fill=white, draw=black]       (dVB4) circle (0.1cm) {}; 

        \node[circle, minimum size=0.5cm,   name=dCB3] at       (dCB.-55) {}; 
        \node[circle, minimum size=0.5cm,   name=dCB4] at       (dCB.-39) {}; 
        \draw[->, >=stealth]                    (dCB3.south) -- (dCB3.north);
        \draw[->, >=stealth]                    (dCB4.north) -- (dCB4.south);
        \filldraw[fill=black, draw=black]       (dCB3) circle (0.1cm) {}; 
        \filldraw[fill=black, draw=black]       (dCB4) circle (0.1cm) {};

        \node[circle, minimum size=0.5cm,   name=eVB1] at       (eVB.105) {}; 
        \node[circle, minimum size=0.5cm,   name=eVB2] at       (eVB.75) {}; 
        \draw[<-, >=stealth]                    (eVB1.south) -- (eVB1.north);
        \draw[<-, >=stealth]                    (eVB2.north) -- (eVB2.south);
        \filldraw[fill=white, draw=black]       (eVB1) circle (0.1cm) {}; 
        \filldraw[fill=white, draw=black]       (eVB2) circle (0.1cm) {}; 

        \node[circle, minimum size=0.5cm,   name=eCB1] at       (eCB.-105) {}; 
        \node[circle, minimum size=0.5cm,   name=eCB2] at       (eCB.-75) {}; 
        \draw[->, >=stealth]                    (eCB1.south) -- (eCB1.north);
        \draw[->, >=stealth]                    (eCB2.north) -- (eCB2.south);
        \filldraw[fill=black, draw=black]       (eCB1) circle (0.1cm) {}; 
        \filldraw[fill=black, draw=black]       (eCB2) circle (0.1cm) {}; 
        
    \end{tikzpicture}
    \caption[Excitonic Quasi-Particles]{Different spin configurations for excitonic quasi-particles in the s-shell:\\
        a) Excitons $\mathrm{X}$ \tab 
        b) Dark excitons $\mathrm{DX}$ \tab 
        c) Positive trions $\mathrm{X}^+$ \tab 
        d) Negative trions $\mathrm{X}^-$ \tab 
        e) Biexciton $\mathrm{XX}$ \\
        The arrows indicate spin direction, while an empty circle depicts a $\mathrm{h}^+$ and a filled circle an $\mathrm{e}^-$. Adapted from \cite[p.~18]{ThesisOpticalCharacterisationOfTelecommunication-WavelengthQDs}.
     } \label{fig:ExcitonicQuasi-Particles}
\end{figure}

\subsection{Entangled Photon Pair Generation via Biexciton-Exciton Cascade}\label{subsec:EntangledPhotonPairGeneration}

The $\mathrm{XX}$ decays via the so-called biexciton-exciton cascade. That is, the biexciton state $\ketXX$ decays into one of two (in the perfect case degenerate) exciton states $\ketXone$ and $\ketXtwo$, which subsequently decay to the ground state of the empty \gls{QD} $\ketG$. This can be seen in Figure \ref{fig:XX-X-Cascade} on the left. The more realistic version of the cascade is shown in Figure \ref{fig:XX-X-Cascade} on the right. First we will discuss the mechanism of the simplified case and then how the more realistic description affects the entangled photon generation. Then follows an explanation why the advanced version of the model is needed to depict reality more accurately.

\begin{figure}
    \centering
        \begin{tikzpicture}[
            scale = 0.5,
            every node/.style={scale=0.75},
            ELvl/.style={black, rectangle, minimum width=1.2cm, minimum height=0.2cm}
        ]
        
        \draw[->] (-1.5, -0.5) -- (-1.5, 8.5) node[above] {$E$ [\qty{}{\joule}]}; 
        
        \node[ELvl, name=G] at (2,0) {};
        \node[ELvl, name=X1] at (0,4) {};
        \node[ELvl, name=X2] at (4,4) {};
        \node[ELvl, name=XX] at (2,8) {};

        \draw (G.west) -- (G.east) node[right] {$\ketG$};
        \draw (X1.west) -- (X1.east) node[right] {$\ketXone$};
        \draw (X2.west) -- (X2.east) node[right] {$\ketXtwo$};
        \draw (XX.west) -- (XX.east) node[right] {$\ketXX$};
        
        \draw[<->, >=stealth, blue] (G.160) -- node[midway, left] {$\lambda_0$}  node[midway, right] {$\ket{\mathrm{L}}$} (X1.south);
        \draw[<->, >=stealth, blue] (G.20) -- node[midway, right] {$\lambda_0$} node[midway, left] {$\ket{\mathrm{R}}$} (X2.south);
        \draw[<->, >=stealth, blue] (X1.north) -- node[midway, left] {$\lambda_0$} node[midway, right] {$\ket{\mathrm{R}}$} (XX.200);
        \draw[<->, >=stealth, blue] (X2.north) -- node[midway, right] {$\lambda_0$}  node[midway, left] {$\ket{\mathrm{L}}$} (XX.340);
        
        \begin{scope}[xshift=8cm] 
        
            \node[ELvl, name=G] at (2,0) {};
            \node[ELvl, name=X1] at (0,4.3) {};
            \node[ELvl, name=X1FSS] at (4, 4.3) {};
            \node[ELvl, name=X2] at (4,3.7) {};
            \node[ELvl, name=X2FSS] at (0, 3.7) {};
            \node[ELvl, name=XX] at (2,7) {};
            \node[ELvl, name=XXEB] at (2,8) {};
    
            \draw (G.west) -- (G.east) node[right] {$\ketG$};
            \draw (X1.west) -- (X1.east) node[above right] {$\ketXone$};
            \draw[dashed, gray] (X1.east) -- (X1FSS.east);
            \draw (X2.west) -- (X2.east) node[below right] {$\ketXtwo$};
            \draw[dashed, gray] (X2FSS.west) -- (X2.west);
            \draw (XX.west) -- (XX.east) node[right] {$\ketXX$};
            \draw[dashed, gray] (XXEB.west) -- (XXEB.east) node[right] {$2 \cdot (E_{\mathrm{X}} - E_{\mathrm{G}})$};
            
            \draw[<->, >=stealth, red] (X1.north) -- node[midway, left] {$\lambda_1$} node[midway, right] {$\ket{\mathrm{V}}$} (XX.200);
            \draw[<->, >=stealth, red] (X2.north) -- node[midway, right] {$\lambda_2$} node[midway, left] {$\ket{\mathrm{H}}$} (XX.340);
            \draw[<->, >=stealth, red] (G.160) -- node[midway, left] {$\lambda_3$} node[midway, right] {$\ket{\mathrm{V}}$} (X1.south);
            \draw[<->, >=stealth, red] (G.20) -- node[midway, right] {$\lambda_4$} node[midway, left] {$\ket{\mathrm{H}}$} (X2.south);

            \draw[<->, >=stealth, gray] (XX.center) -- node[left] {$E_\text{b}$} (XXEB.center);
            \draw[<->, >=stealth, gray] (X2.east) -- node[right] {$\Delta$} (X1FSS.east);
            
        \end{scope}
    \end{tikzpicture}
    \caption{In the idealised biexciton-exciton cascade (left), exciton levels are degenerate, with uniform energy differences between the ground state $\ketG$ and exciton states $\ketXone$ or $\ketXtwo$, as well as between the biexciton state $\ketXX$ and the exciton states. In contrast, the non-idealised biexciton-exciton cascade (right) features distinct energy differences for each level transition, characterized by unique wavelengths. The degeneracy of the exciton levels is lifted by the Fine Structure Splitting (FSS) $\Delta$. The binding energy between two excitons is denoted $E_\text{b}$, the energy of the ground level is denoted $E_{\mathrm{G}}$, and the energy of the idealised exciton level is denoted $E_{\mathrm{X}}$.}
    \label{fig:XX-X-Cascade}
\end{figure}

\subsubsection{Idealised Biexciton-Exciton Cascade}\label{subsubsec:XX-X-CascadeIdealised}

As signified in the figure the decay produces a pair of photons in the following way: Due to the considerations about conservation of angular momentum above, the transitions couple to circularly polarised light. Thus the decay from the $\mathrm{XX}$ to the $\mathrm{X}$ produces 
right circularly polarised light when the resulting state is $\ketXone$ and left circularly polarised light when the resulting state is $\ketXtwo$.
This produces a photon which is entangled with the state of the \gls{QD} (\textit{spin-photon entanglement}). The subsequent decay of the $\mathrm{X}$ to the empty \gls{QD} produces another photon which then is entangled with the first photon. Depending on the excitonic state present in the \gls{QD} after the first decay the second decay produces right or left circularly polarised light 
if the first released photon was left or right circularly polarised, respectively. Because of the degenerate exciton energy levels, the decay paths are indistinguishable and coexist in superposition. In total this yields the following entangled state of the released photons \cite[pp.17 et seqq.]{DissZeuner}:
\begin{equation}
    \ket{\Psi} = \frac{1}{\sqrt{2}}(\ket{\mathrm{RL}} + \ket{\mathrm{LR}}) = \frac{1}{\sqrt{2}}(\ket{\mathrm{HH}} + \ket{\mathrm{VV}}).
\end{equation}

\subsubsection{Biexciton-Exciton Cascade with Improved Accuracy}\label{subsubsec:XX-XCascadeWithImprovedAccuracy}

In the more realistic case shown on the right in Figure \ref{fig:XX-X-Cascade} the exciton levels are not degenerate due to a finite \gls{FSS} (for the reason for its existence and its calculation see Section \ref{section:Physics}.\ref{subsubsec:ModelAccuracy}) denoted $\Delta$ breaking the indistinguishability of the decay paths yielding following time-dependent state of the released photons:
\begin{equation}\label{eq:time-dependent_photon_release}
    \ket{\Psi(t)} = \frac{1}{\sqrt{2}}\left(\ket{\mathrm{HH}} + \exp\left(\frac{i t \Delta}{\hbar}\right)\ket{\mathrm{VV}}\right).
\end{equation}
This means there is an oscillation in time $t$ of the exciton between the two eigenstates $\ketXone$ and $\ketXtwo$ after emission of the first photon.
Thus, after the second emission there is an oscillation between the Bell states 
\begin{equation}
    \frac{1}{\sqrt{2}}(\ket{\mathrm{HH}} + \ket{\mathrm{VV}})
\end{equation}
and
\begin{equation}
    \frac{1}{\sqrt{2}}(\ket{\mathrm{HH}} - \ket{\mathrm{VV}})
\end{equation}
dependent on the detection time of the second photon in relation to the emission of the first one. Even with finite \gls{FSS} the photons are still in the maximally entangled state, but one has to have a high enough resolution in the respective experimental setup to resolve the oscillation with time constant $\frac{\hbar}{\Delta}$.
\cite[pp.17 et seqq.]{DissZeuner}

Additionally, the energy level of the biexciton is also not the same energy as the energy of two separate excitons, because it is shifted by the binding energy $E_\text{b}$ (see Figure \ref{fig:XX-X-Cascade}), which is also explained in Section \ref{section:Physics}.\ref{subsubsec:ModelAccuracy}.
\cite{DissSimonGordon}

\subsubsection{Model Accuracy}\label{subsubsec:ModelAccuracy}

As alluded to before, there are physical phenomena giving rise to inaccuracies in the idealised biexciton-exciton cascade model. Some of those phenomena were incorporated into the more advanced version of the model to improve the accuracy and they will be discussed here now.

In a simple single-particle model the resulting energy structure consists of four energy states namely the transitions of 
spin $+\frac{1}{2}$ $\mathrm{e}^-$ and angular momentum $+\frac{3}{2}$ $\mathrm{h}^+$,
spin $-\frac{1}{2}$ $\mathrm{e}^-$ and angular momentum $-\frac{3}{2}$ $\mathrm{h}^+$,
spin $+\frac{1}{2}$ $\mathrm{e}^-$ and angular momentum $-\frac{3}{2}$ $\mathrm{h}^+$, and 
spin $-\frac{1}{2}$ $\mathrm{e}^-$ and angular momentum $+\frac{3}{2}$ $\mathrm{h}^+$ the latter two of which are optically allowed.
\cite[p.16]{DissSimonGordon}

But as soon as there are more than one particle in the \gls{QD} the single-particle model is not valid anymore.
The multi-particle model used for the description of the \gls{QD} with multiple particles (\eg an $\mathrm{e}^-$ and an $\mathrm{h}^+$) also takes into account the interactions between the particles which leads to renormalisation of energies. Thus, if there are one $\mathrm{h}^+$ and $\mathrm{e}^-$ each in the \gls{QD}: due to the Coulomb interaction between them they constitute an electron-hole pair (\ie the exciton) whose energy is reduced in comparison to the sum of energies of the separate particles from the single-particle model. They are grouped into the different categories described in Section \ref{section:Physics}.\ref{subsec:ExcitonicQuasi-Particles} and Figure \ref{fig:ExcitonicQuasi-Particles}.
By just taking the Coulomb interaction into account all four energy levels of the $\mathrm{DX}$ and the $\mathrm{X}$ are degenerate. 

The localisation of charge carriers in the small volume of the \gls{QD} warrants the consideration of exchange interaction as well. This firstly lifts the degeneracy of the energy levels of dark and bright excitons and also mixes the energy states of the dark excitons into two separate ones (\textit{hybridisation}), resulting in two degenerate bright exciton states which are separate from two non-degenerate dark exciton states (which will not be considered in our modelling here as they are not optically active).
\cite[pp.16-18]{DissSimonGordon}

Furthermore, as mentioned in Section \ref{section:Physics}.\ref{subsubsec:XX-XCascadeWithImprovedAccuracy} there is the \gls{FSS} $\Delta$. The idealised version of the \gls{QD} used for the model in Figure \ref{fig:XX-X-Cascade} on the left assumes rotational symmetry of the \gls{QD} in the plane perpendicular to its growth direction (point group $D_{2d}$ in \textit{Schoenflies notation} \cite{Schoenflies}). In case of a "more asymmetric" \gls{QD} (group $C_{2v}$ or $C_2$) one has a \gls{FSS} due to the mixing of the two exciton states which also results in the coupling of these energy levels to linear polarised light (mixture of both circular polarizations). Regarding the exciton states there are additional effects like weak coupling of $\mathrm{DX}$ to light in growth direction of the \gls{QD} and mixing of all four exciton states due to completely asymmetric \glspl{QD} which will not be discussed here or incorporated into our models.
\cite[pp.17-18]{DissSimonGordon}

The value for the \gls{FSS} can be calculated as follows:
\begin{align}
\begin{split}
    \Delta  &= (E_{\mathrm{XX}} - E_{\mathrm{X}_1}) - (E_{\mathrm{XX}} - E_{\mathrm{X}_2}) \\
            &= E_{\mathrm{V,XX}} - E_{\mathrm{H,XX}}
\end{split}
\end{align}
or equivalently:
\begin{align}
\begin{split}
    \Delta  &= (E_{\mathrm{X}_2} - E_{\mathrm{G}}) - (E_{\mathrm{X}_1} - E_{\mathrm{G}}) \\ 
            &= E_{\mathrm{H,X}} - E_{\mathrm{V,X}},
\end{split}
\end{align}
where 
$E_{\mathrm{X}_1}$ is the energy of the exciton coupling to vertically polarised light, 
$E_{\mathrm{X}_2}$ is the energy of the exciton coupling to horizontally polarised light. Thus recalling Figure \ref{fig:XX-X-Cascade} we can shorten this, where
$E_{\mathrm{V,XX}}$ is the energy corresponding to $\lambda_{1}$,
$E_{\mathrm{H,XX}}$ to $\lambda_{2}$,
$E_{\mathrm{V,X}}$ to $\lambda_{3}$, and
$E_{\mathrm{H,X}}$ to $\lambda_{4}$.

In principle, either calculating the \gls{FSS} via the transitions from the biexciton state to the exciton states or via the transitions from the exciton states to the ground state should yield the same absolute value for $\Delta$. 
But practically speaking the measurements of energies are subject to noise and possibly systematic errors. Thus one can apply the following way of calculating $\Delta$ to minimise systematic errors:
\begin{equation}
    \Delta = \frac{(E_{\mathrm{H,X}} - E_{\mathrm{V,X}}) + (E_{\mathrm{V,XX}} - E_{\mathrm{H,XX}})}{2},
\end{equation}
implying a positive \gls{FSS} if the exciton coupling to horizontal polarization has higher energy.
\cite{InversionXLevelSplittinInQDs_FSSFormula}

Now, for the binding energy $E_\text{b}$ which shifts the actual biexciton level w.r.t.~the level of two separate non-interacting excitons (denoted as $2 \cdot (E_{\mathrm{X}} - E_{\mathrm{G}})$ in Figure \ref{fig:XX-X-Cascade}, where $E_{\mathrm{G}}$ is the ground level energy and $E_{\mathrm{X}}$ the idealised exciton level without \gls{FSS}). Because there is some interaction between the two excitons when they are in the same \gls{QD} the energy level of the biexciton is shifted. If the transition energy between biexciton and single exciton is smaller than the transition energy from the exciton to the ground state the biexciton is in the so called binding state, otherwise it is in the anti-binding state. The actual value for $E_\text{b}$ is highly dependent on the material system, \eg around 
\qty{43,259E-21}{\joule} 
(\qty{2,7}{\milli\electronvolt})
for \ch{InAs}/\ch{GaAs} \glspl{QD}.
\cite[pp.18-19]{DissSimonGordon}

\subsubsection{Excitation Schemes}\label{subsubsec:ExcitationSchemes}

There are different schemes to excite the \gls{QD} and here we provide a very short introduction to some of them. This should by no means be viewed as an exhaustive review of excitation schemes.

The exciton transition can \eg be optically excited nonresonantly, resonantly or with phonon-assisted excitation. With resonant excitation no free charge carriers are introduced reducing electronic noise, which yields near transform-limited linewidths. It can also produce high indistinguishability, but single-photon purity is limited to $g^{(2)}(0) \approx 10^{-2}$ by re-excitation and then release of another photon.
\cite{GenerationIndistinguishableSinglePhotons}

For exciting a \gls{QD} in the ground state to the biexciton level one can \eg use the resonant \gls{TPE} scheme \cite{ResonantTwoPhotonExcitation} or detuned phonon-mediated excitation \cite{DissipativePrepOfXAndXXInQDs}.~\cite{GenerationNonClassicalLightWithSemiconductorQDs}
Due to the conservation of angular momentum it makes sense that the transition \textit{directly} from the ground state to the biexciton does not couple to a single photon (angular momentum $\pm 1$) but to two photons via a virtual exciton as the biexciton consists of two coupled excitons with total spin $0$.~\cite{PhononAssistedXXGenerationInQD}
If one uses \gls{TPE} of the biexciton one is able to suppress re-excitation in single-photon generation leading ultralow multiphoton errors, albeit at the cost of limited indistinguishability due to the cascade.~\cite{GenerationIndistinguishableSinglePhotons}

Resonant \gls{TPE} has also been practically demonstrated for entangled photon pair generation \cite{QuantumDotEntangledPhotonPairSource} and the simple model we employ here uses this mechanism in a simplified way by just employing one single interaction wavelength $\lambda_0$ as seen in Figure \ref{fig:XX-X-Cascade}. In the more accurate model resonant \gls{TPE} of the biexciton is incorporated as follows.

\paragraph{Resonant Two-Photon Excitation of the Biexciton}

In resonant excitation one excites the energy level by introducing light to the system which has the same amount of energy as the transition to be excited. In case of the resonant \gls{TPE} one uses two photons to get from $\ketG$ to $\ketXX$ both of which have half the energy of this transition, \ie both of them in conjunction have the correct amount of energy.
In more detail, the excitation is done with a pulsed laser which has energy $E_{\text{exc}} = (E_{\mathrm{XX}} - E_{\mathrm{G}})/2$ with linear polarization $\ket{\mathrm{H}}$, where $E_{\mathrm{XX}}$ is the energy corresponding to the state $\ketXX$ and $E_{\mathrm{G}}$ corresponding to $\ketG$. The laser's light couples to the ground state $\ketG$ and the biexciton state $\ketXX$ via a virtual level (see dashed line labelled $E_{\text{exc}}$ in Figure \ref{fig:ResonantTwo-PhotonExcitation}). Afterwards, the $\mathrm{XX}$ can decay as discussed before. Due to the biexciton binding energy $E_\text{b}$  one has:
\begin{equation}
    E_{\mathrm{X}_{\{1,2\}}} - E_{\mathrm{G}} > E_{\text{exc}} > E_{\mathrm{XX}} - E_{\mathrm{X}_{\{1,2\}}}
\end{equation}
with the energies of either exciton labelled $E_{\mathrm{X}_{\{1,2\}}}$.
\cite{ResonantTwoPhotonExcitation}

Thus, an additional excitation wavelength $\lambda_5$ is introduced in the more accurate model (see Figure \ref{fig:ResonantTwo-PhotonExcitation}), while in the simplified model $\lambda_0$ is utilised to model both excitation and emission from the energy structure (Figure \ref{fig:XX-X-Cascade}).

\begin{figure}
    \centering
        \begin{tikzpicture}[
            scale = 0.75,
            every node/.style={scale=0.75},
            ELvl/.style={black, rectangle, minimum width=1.2cm, minimum height=0.2cm}
        ]

        \draw[->] (-1.5, -0.5) -- (-1.5, 8.5) node[above] {$E$ [\qty{}{\joule}]}; 
        
        \node[ELvl, name=G] at (2,0) {};
        \node[ELvl, name=X1] at (0,4.2) {};
        \node[ELvl, name=X1FSS] at (4, 4.2) {};
        \node[ELvl, name=X2] at (4,3.8) {};
        \node[ELvl, name=X2FSS] at (0, 3.8) {};
        \node[ELvl, name=XX] at (2,6.5) {};
        \node[ELvl, name=XXEB] at (2,8) {};

        \node[ELvl, name=EexcL] at (0, 3.25) {};
        \node[ELvl, name=EexcM] at (2, 3.25) {};
        \node[ELvl, name=EexcR] at (4, 3.25) {};
    
        \draw (G.west) -- (G.east) node[right] {$\ketG$};
        \draw (X1.west) node[left] {$\ketXone$} -- (X1.east);
        \draw (X2.west) -- (X2.east) node[right] {$\ketXtwo$};
        \draw (XX.west) -- (XX.east) node[right] {$\ketXX$};
        \draw[dashed, gray] (XXEB.west) -- (XXEB.east) node[right] {$2 \cdot (E_{\mathrm{X}} - E_{\mathrm{G}})$};

        \draw[dashed, gray] (EexcL.west) -- (EexcR.east) node[right] {$E_{\text{exc}}$};
            
        \draw[<-, >=stealth, red] (X1.north) -- node[midway, left] {$\lambda_1$} node[midway, right] {$\ket{\mathrm{V}}$} (XX.200);
        \draw[<-, >=stealth, red] (X2.north) -- node[midway, right] {$\lambda_2$} node[midway, left] {$\ket{\mathrm{H}}$} (XX.340);
        \draw[<-, >=stealth, red] (G.160) -- node[midway, left] {$\lambda_3$} node[midway, right] {$\ket{\mathrm{V}}$} (X1.south);
        \draw[<-, >=stealth, red] (G.20) -- node[midway, right] {$\lambda_4$} node[midway, left] {$\ket{\mathrm{H}}$} (X2.south);

        \draw[->, >=stealth, cyan] (G.north) -- node[right, pos=0.4] {$\lambda_5$} node[left, pos=0.4] {$\ket{\mathrm{H}}$} (EexcM.south);
        
        \draw[->, >=stealth, cyan] (EexcM.north) -- node[right, pos=0.4] {$\lambda_5$} node[left, pos=0.4] {$\ket{\mathrm{H}}$} (XX.south);

        \draw[<->, >=stealth, gray] (XX.center) -- node[left] {$E_\text{b}$} (XXEB.center);

    \end{tikzpicture}
    \caption{The mechanism of resonant \gls{TPE} of the biexciton. The excitation laser is marked in cyan.}
    \label{fig:ResonantTwo-PhotonExcitation}
\end{figure}

\paragraph{Detuned Phonon-Mediated Excitation of the Biexciton}

The biexciton can also be excited via a \gls{TPE} which has an energy which is detuned from the biexciton level by some amount (this can be seen as the energy denoted $E_{\Delta_{\text{2photon}\mathrm{XX}}}$ in Figure \ref{fig:Phonon-AssistedTwo-PhotonExcitation}). That means the two photons meant to excite from the ground state to the biexciton have each the energy $E_{\text{exc}} = (E_{\Delta_{\text{2photon}\mathrm{XX}}} - E_{\mathrm{G}})/2$ with associated wavelength $\lambda_4$. The excess energy of the detuned excitation pulse decays nonradiatively to the $\ketXX$ state mediated by phonons. According to \citeauthor{DissipativePrepOfXAndXXInQDs} in \cite{DissipativePrepOfXAndXXInQDs} this excitation scheme is deterministic, fast, and provides a high fidelity, while it does not require such a high precision in the control of the excitation power as coherent \gls{TPE} in resonance, making it more robust to decoherence. After this biexciton preparation the cascade can be utilised to generate polarization entangled photons.
\cite{DissipativePrepOfXAndXXInQDs}

The effect of different detuning values and their influence on the biexciton population can be found in \cite{DissipativePrepOfXAndXXInQDs}. But in principle different values are possible which might shift the level of $E_{\text{exc}}$ accordingly.

Even though it was expected that phonon-assisted generation of the biexciton could be a major source of decoherence \cite{PhononAssistedXXGenerationInQD}, \citeauthor{PhononAssistedRobustDeterministic2PhotonXXPreparationInQD} show that photons generated with this scheme can exhibit similar coherence times to resonant \gls{TPE} \cite{PhononAssistedRobustDeterministic2PhotonXXPreparationInQD}.

\begin{figure}
    \centering
        \begin{tikzpicture}[
            scale = 0.75,
            every node/.style={scale=0.75},
            ELvl/.style={black, rectangle, minimum width=1.2cm, minimum height=0.2cm}
        ]

        \draw[->] (-1.5, -0.5) -- (-1.5, 8.5) node[above] {$E$ [\qty{}{\joule}]}; 
        
        \node[ELvl, name=G] at (2,0) {};
        \node[ELvl, name=X1] at (0,4.2) {};
        \node[ELvl, name=X1FSS] at (4, 4.2) {};
        \node[ELvl, name=X2] at (4,3.8) {};
        \node[ELvl, name=X2FSS] at (0, 3.8) {};
        \node[ELvl, name=XX] at (2,6.5) {};
        \node[ELvl, name=XXEB] at (2,8) {};

        \node[ELvl, name=EexcL] at (0, 4.00) {};
        \node[ELvl, name=EexcM] at (2, 4.00) {};
        \node[ELvl, name=EexcR] at (4, 4.00) {};

        \node[ELvl, name=D2PXX] at (2, 8) {};
    
        \draw (G.west) -- (G.east) node[right] {$\ketG$};
        \draw (X1.west) node[left] {$\ketXone$} -- (X1.east);
        \draw (X2.west) -- (X2.east) node[below right] {$\ketXtwo$};
        \draw (XX.west) -- (XX.east) node[right] {$\ketXX$};

        \draw[dashed, gray] (EexcL.west) -- (EexcR.east) node[right] {$E_{\text{exc}}$};

        \draw[dashed, gray] (D2PXX.west) -- (D2PXX.east) node[right] {$E_{\Delta_{\text{2photon}\mathrm{XX}}}$};
            
        \draw[<-, >=stealth, red] (X1.north) -- node[midway, left] {$\lambda_1$} node[midway, right] {$\ket{\mathrm{V}}$} (XX.200);
        \draw[<-, >=stealth, red] (X2.north) -- node[midway, right] {$\lambda_2$} node[midway, left] {$\ket{\mathrm{H}}$} (XX.340);
        \draw[<-, >=stealth, red] (G.160) -- node[midway, left] {$\lambda_3$} node[midway, right] {$\ket{\mathrm{V}}$} (X1.south);
        \draw[<-, >=stealth, red] (G.20) -- node[midway, right] {$\lambda_4$} node[midway, left] {$\ket{\mathrm{H}}$} (X2.south);

        \draw[->, >=stealth, cyan] (G.north) -- node[right, pos=0.7] {$\lambda_5$} (EexcM.south);
        
        \draw[->, >=stealth, cyan] (EexcM.north) -- node[right, pos=0.2] {$\lambda_5$} (D2PXX.south);

        \draw[->, >=stealth, gray, dashed] (D2PXX.340) -- node[right, pos=0.5] {nonradiative decay} (XX.20);

    \end{tikzpicture}
    \caption{The mechanism of detuned phonon-assisted \gls{TPE} of the biexciton. The excitation laser is marked in cyan.}
    \label{fig:Phonon-AssistedTwo-PhotonExcitation}
\end{figure}

\paragraph{Adiabatic Rapid Passage Chirped Excitation}

More advanced schmes for excitation include \gls{ARP} excitation.
Generally speaking, \gls{ARP} (frequency-)chirped excitation is the excitation of a transition with a laser pulse whose frequency is swept through the resonance frequency of the transition.~\cite{QDExcitonGenerationViaAdiabaticPassage} The change in frequency has to be slow enough for the transition to happen adiabatically. 
This can be done either with positively chirped pulses (frequency swept through the resonance frequencies from low to high) or with a negative chirp (the other way around).

It was shown for single photon generation in a \gls{QD} via \gls{ARP} that the population transfer between the states is quite insensitive to interaction time, laser power fluctuation, and pulse area when compared to resonance fluorescence with transform-limited $\pi$-pulses. 
\gls{ARP} with positive chirp is even less sensitive to pulse area than negative chirp. Additionally, \gls{ARP} may exhibit linewidths which can be compared to those observed when driving with transform-limited $\pi$-pulses.
\cite{DeterministicGenerationOfSinglePhotonsQDUsingARP}

\gls{ARP} chirped excitation can also be used for entangled photon generation. Yet, excitation up to the biexciton level is only robust for large enough positive chirps and low temperatures due to acoustic phonons. This is then insensitive to small variations of pulse area, but for biexciton generation \gls{ARP} is sensitive to the sign of the chirp even without modelling carrier-phonon interaction. To achieve population transfer to the biexciton from the ground state one can use a linearly polarised, chirped Gaussian pulse with a central frequency in resonance with half of the transition energy from $\ketG$ to $\ketXX$, \ie for zero chirp it coincides with resonant \gls{TPE}.
For negative chirps robust biexciton generation is only possible if the biexciton binding energy $E_\text{b}$ is either very small or large, which is one of the important differences to \gls{ARP} for single exciton generation.
\cite{BiexcitonInQDviaARP}

\paragraph{Two-Colour Adiabatic Rapid Passage Chirped Excitation}

If two different wavelengths are used for \gls{ARP} excitation it is called two-colour \gls{ARP}. That is, it is possible to use two circularly polarised chirped Gaussian pulses to excite the \gls{QD} with central frequencies resonant to the transition from $\ketG$ to $\ket{\mathrm{X}}$ and from $\ket{\mathrm{X}}$ to $\ketXX$, for the two pulses respectively. These pulses of equal chirp are applied at the same time. This scheme is also robust to changes in pulse intensity and chirp, when ignoring the exciton-phonon coupling (as long as the conditions for \gls{ARP} are still met). A fixed time delay between the two different pulses also does not impact the efficiency of the scheme significantly. The only condition in this case of adiabatic evolution is the ordering, \ie the pulse exciting $\ketG$ to $\ket{\mathrm{X}}$ is before the pulse exciting $\ket{\mathrm{X}}$ to $\ketXX$ or both pulses have enough temporal overlap. The scheme is only robust for positive chirps just like the monochrome \gls{ARP} for biexciton generation, if one considers carrier-phonon coupling.
\cite{BiexcitonInQDviaARP}

\paragraph{Dichromatic Pulsed Excitation}\label{sec:phys-dichromatic-excitation}

Two-colour excitation or \gls{DPE} is an excitation scheme to coherently drive a two-level transition -- \eg in \glspl{QD} -- that utilises two pulses with a given envelope, which are both detuned in frequency by $\pm \Delta_{\text{DPE}}$ from the given transition. Conceptually, the combination of both identical, equally detuned pulses should become equivalent to a single pulse that is resonant, yet with a modified envelope. The advantage of this scheme should be the spectral separation of the excitation pulses and the targeted transition, making it easy to filter out the excitation pulses. Nevertheless, there is no excited state population in the two-level system after the excitation pulse, except if parts of the dichromatic excitation spectrum overlap with the target transition. For the excitation with equal excitation pulses (symmetric \gls{DPE}) of a dissipation-free two-level system the temporal pulse area of the excitation cancels out in total. Thus, also the excited state population vanishes. When dissipation or additional interactions are taken into account this can lead to not completely cancelling pulse area, in turn leading to nonzero excited state population. 
And this is also possible if there is no spectral overlap of a \gls{DPE} pulse with the transition. Larger excited state populations can be achieved with \gls{DPE} pulses of different intensity (asymmetric \gls{DPE}).
\cite{DichromaticPulsedExcitationDPE}

To excite a $\mathrm{XX}$-$\mathrm{X}$-cascade system one can use equal \gls{DPE} pulses which are detuned by $\pm \Delta_{\text{DPE}}$ from the $\ket{\mathrm{X}}$ transition. Under resonant \gls{TPE} one would get Rabi oscillations from both the biexciton and exciton emissions. In case one of the pulses for \gls{DPE} overlaps with the energy of resonant \gls{TPE} (which was denoted $E_{\text{exc}}$ in Figure \ref{fig:ResonantTwo-PhotonExcitation}), Rabi oscillations are also visible which vanish if there is minimal overlap with $E_{\text{exc}}$.
\cite{DichromaticPulsedExcitationDPE}

\paragraph{Swing-Up of Quantum Emitter Population}

Swing-up of quantum emitter population -- swing-up scheme for short -- is yet another coherent excitation scheme for a target transition with nonresonant pulses, which can thus be filtered out easily and does not rely on incoherent mediators such as phonons. Because its pulses are highly detuned they do not lead to significant population transfer to the excited state individually, yet by periodically changing which pulse is applied the population is gradually transferred with a swing-up behaviour. To achieve this, the total excitation pulse can be modulated discretely, \ie switching between intervals where one of the two different detunings (denoted $\Delta_{\text{low}}$ for the smaller and $\Delta_{\text{high}}$ for the bigger detuning) are applied. For each detuning individually the Rabi oscillation has a very small amplitude, yielding negligible population in the excited state when they would be applied separately. For every increase in population of the excited state $\Delta_{\text{low}}$ with the higher Rabi oscillation amplitude is used and for every decrease in population $\Delta_{\text{high}}$ with the lower Rabi oscillation amplitude is used, gradually escalating the excited state population, leading to the eponymous "swing-up" of population. A sinusoidal frequency modulation between the two detunings could in principle also be used for the swing-up. 
Experimentally, such a frequency modulation might be hard to achieve, so it is also possible to apply a two-colour variant of the scheme. That is, because the swing-up scheme relies on the usage of different Rabi oscillations, one can use amplitude modulation to change the Rabi frequency. This in turn can be achieved by superimposing two Gaussian pulses with similar width but two different detunings (hence two colours). 
In general, the swing-up scheme works even if none of the spectral components of the excitation pulse are resonant with the transition to be driven. Additionally, it is possible to use pulses which are both detuned below the transition energy, thus using higher pulse areas is less problematic than \eg with \gls{DPE}.
\cite{Swing-UpOfQuantumEmitterPopulationUsingDetunedPulses}

Excitation of a biexciton in the cascade multi-level system via the swing-up scheme is also possible. To that end, two distinct pulses are needed. The first one is detuned lower than the resonant \gls{TPE} excitation energy ($E_{\text{exc}}$ in Figure \ref{fig:ResonantTwo-PhotonExcitation}), such that it is not resonant with any exciton transition. The dressed state analysis of such a multi-level system does not yield clear criteria for the second pulse to excite the biexciton to a high degree directly. Yet, through numerical parameter search to construct the second pulse one can find parameters that lead to proper biexciton population after both pulses. During such pulses both the exciton and the biexciton are addressed, but after the pulses only the biexciton is populated.
\cite{Dressed-stateAnalysisOfTwo-ColorExcitationSchemes}

\paragraph{Stimulated Emission from the Biexciton}

For completeness we also mention what would happen if the recombination of the biexciton is triggered by stimulated emission. If the emission from the biexciton is stimulated with accurately timed pulses the timing jitter -- and thus indistinguishability of the resulting photon -- is very low. But the polarization of the stimulating light also determines the polarization of the resulting photon.\footnote{That means in our more accurate model: A stimulating $\ket{\mathrm{V}}$ ($\ket{\mathrm{H}}$) pulse will result in emission of $\ket{\mathrm{V}}$ ($\ket{\mathrm{H}}$) from the transition from $\ketXX$ to $\ket{\mathrm{X}}$}~\cite{GenerationIndistinguishableSinglePhotons} 
The stimulated emission from the biexciton is mostly studied in the context of single photon generation for the $\mathrm{XX}$-$\mathrm{X}$-cascade:
It strictly determines which path of the cascade is taken, which is contrary to the situation for spontaneous emission. If a \gls{TPE} scheme is used for biexciton generation the Bell-like state that is created is then again disentangled by the stimulation pulse.~\cite{StimEmissionFromBiexcitonInQD}
Additionally, it is experimentally problematic to distinguish between the stimulating pulse and the first emitted photon of the cascade.
In conclusion, stimulated emission obviously is advantageous if one wants to control the polarization of the emission for deterministic single photon generation. But, due to aforementioned disentangling by the stimulation pulse this is opposed to the main focus of this work, which is the generation of entangled photon pairs.

\subsection{Realistic Experimental Parameter Sets}

Thus, it is possible to use the $\mathrm{XX}$-$\mathrm{X}$-cascade for entangled photon pair generation. This is routinely performed experimentally (\eg \cite{QuantumDotEntangledPhotonPairSource}), and in order to at least give an idea for a realistic parameter set we provide some examples here. This cannot replace a thorough review of experimental realisations, but for that we refer the reader to other literature.
Especially recent work towards \glspl{QD} emitting at telecommunication wavelengths, \eg by tuning the emission wavelength of \ch{GaAs} \glspl{QD} towards the O-band around \qty{1310}{\nano\meter} and the C-band around \qty{1550}{\nano\meter} with a graded \ch{In_{x}Ga_{1-x}As} metamorphic buffer layer \cite{Scaparra2023} is relevant for applications in real networks.

Experimentally one might encounter also trionic and other multi-exciton states apart from the biexciton, which might exhibit significantly more complex energy structures whose complexities are not easily captured \cite{DissSimonGordon}.
For example, even when utilising the resonant \gls{TPE} mechanism there can be parasitic excitation of trionic states in case $E_{\text{exc}}$ of the laser is too close to said trionic transition \cite{QuantumDotEntangledPhotonPairSource}.
But because those unwanted excitations are not really relevant for entangled photon generation and their effects can be sufficiently small, those effects are neglected in our model here. 

An exemplary set of parameters for the cascade could be \eg from Molecular Beam Epitaxy (MBE) grown \ch{InAs} \glspl{QD} on a graded \ch{InGaAs} buffer layer emitting in the C-Band: \citeauthor{Scaparra2024} report emissions with
$E_{\mathrm{XX}} \approx \qty{803}{\milli\electronvolt}$ corresponding to $ \sim\qty{1544,02}{\nano\meter}$,
$E_{\mathrm{X}} \approx \qty{805}{\milli\electronvolt}$ corresponding to $ \sim\qty{1540,183}{\nano\meter}$
and \gls{FSS}
$\Delta = \qty{55(6)}{\micro\electronvolt}$
for non-resonant continuous wave excitation with \qty{780}{\nano\meter} at \qty{4}{\kelvin}.
\cite{Scaparra2024} 

Another more standard realisation not in the telecommunication bands is, \eg a droplet etched \ch{GaAs} \gls{QD} in optical antenna structures excited with resonant \gls{TPE} emitting around \qty{780}{\nano\meter}: \citeauthor{QuantumDotEntangledPhotonPairSource} report emissions with
$E_{\mathrm{XX}} \approx \qty{1.5871}{\electronvolt}$ corresponding to $ \sim\qty{781,203}{\nano\meter}$,
$E_{\mathrm{X}} \approx \qty{1,5910}{\electronvolt}$ corresponding to $ \sim\qty{779,288}{\nano\meter}$
and \gls{FSS}
$\Delta = \qty{3,9(0,3)}{\micro\electronvolt}$.
\cite{QuantumDotEntangledPhotonPairSource}

For reference, resonant \gls{TPE} can be achieved experimentally with pulsed laser light \cite{ResonantTwoPhotonExcitation, ResonantTwoPhotonExcitation2, ResonantTwoPhotonExcitation3, QuantumDotEntangledPhotonPairSource} that exhibits the right energy and polarization (see Figure \ref{fig:ResonantTwo-PhotonExcitation}), and pulse shape. 
The excitation process via resonant \gls{TPE} can be modelled via a semi-classical Hamiltonian \cite{ResonantTwoPhotonExcitation, ResonantTwoPhotonExcitation3}, which can yield Rabi oscillations and excitations of different form depending on the pulse \cite{ResonantTwoPhotonExcitation} and damping of these oscillations by relaxation processes. This approach was also adopted in later sections of this work.

With such experimental realisations readily available it is possible to work towards applications like \gls{QKD} based on polarisation entangled photons and other schemes. 

\section{Mathematics}\label{section:Mathematics}

Building on the physical system description provided in Section \ref{section:Physics}, this chapter develops a comprehensive mathematical model for the biexciton-exciton cascade within a \gls{QD} system. This model forms the foundation for the implementation of the simulation module discussed in the Section \ref{section:Simulation}.

\subsection{Quantum Dot Hilbert Space}\label{subsection:qd_hilbert_space}

In the context of analysing the biexciton-exciton cascade within a \gls{QD}, it is crucial to define and understand the structure of the associated Hilbert space \HQD, which is effectively isomorphic to $\mathbb{C}^4$ for our purposes \cite{dey2015biexciton}. This four-dimensional complex vector space is spanned by a set of basis vectors corresponding to the physically distinct states of the \gls{QD}: the ground state $\ketG = [1,0,0,0]^T$, which represents the \gls{QD} without excitations; two exciton states, $\ketXone = [0,1,0,0]^T$ and $\ketXtwo = [0,0,1,0]^T$, reflecting the presence of a single exciton; and the biexciton state $\ketXX = [0,0,0,1]^T$, indicative of two bound excitons within the \gls{QD}. Importantly, as illustrated in Figure \ref{fig:XX-X-Cascade}, the exciton states $\ketXone$  and $\ketXtwo$  are degenerate in an idealized \gls{QD} system, meaning they share the same energy level. This degeneracy occurs under ideal conditions without perturbations like \gls{FSS}, despite their differing properties such as polarization coupling.

To model the state transition of the \gls{QD}, we construct an operator $\sigmaOp{S_1}{S_2}$. This operator transforms the state of the \gls{QD} to state $S_2$ if the original state was $S_1$:
\begin{align}
    \label{eq:sigma}
    \begin{split}
    \sigmaOp{S_1}{S_2} &: \HQD \to \HQD,\\
    \sigmaOp{S_1}{S_2} &= \op{S_2}{S_1},
    \end{split}
\end{align}
where the states $S_1, S_2 \in \{ \mathrm{G}, \mathrm{X_1}, \mathrm{X_2}, \mathrm{XX}\}$ represent any \gls{QD} eigenstate.

\subsection{Light Mode Hilbert Space}

The mathematical framework for modelling the state space of light modes is fundamentally established within the Hilbert space formalism. Specifically, we make use of the Fock space representation to describe the quantum states of light, which allows for a comprehensive account of the quantized nature of the electromagnetic field. In this formalism, the basis states of the Fock space also known as number states, provide natural means to represent states with discrete photon number.

To model the essential physical degrees of freedom, such as polarization, we define a separate Fock space for each polarization mode. This approach aligns with the bosonic \cite{bach1990indistinguishability} nature of photons, where indistinguishability and symmetrization require careful handling of mode labelling.

Consequently, each spatio-temporal mode is uniquely defined by its wavelength $\lambda$ and polarization label $p\in\{\mathrm{H},\mathrm{V}\}$. The Hilbert space of a single pulse, resolved by polarization, is then given by
\begin{equation}
  \mathcal{H}_{\lambda} = \mathcal{F}_{\lambda, \mathrm{H}} \otimes \mathcal{F}_{\lambda, \mathrm{V}}.
\end{equation}

Since the state of the individual spatio-temporal light mode is encoded into two Fock spaces and the coupling of the \gls{QD} to specific polarization is determined by the internal properties of said \gls{QD}, more specifically to rotation $\theta$ of the polarization, we need to define specific ladder operators for our system:
\begin{equation}
\hat{a}_{\lambda, \pm}, \hat{a}^\dagger_{\lambda, \pm}: \mathcal{H}_{\lambda} \to \mathcal{H}_{\lambda}
\end{equation}
\begin{equation}
\begin{aligned}
\hat a^\dagger_{\lambda,+}(\theta) &=~~~ \cos(\theta)\hat a^\dagger_{\lambda, \mathrm{H}} \!&+~ i \sin(\theta)\hat a^\dagger_{\lambda, \mathrm{V}}, \\
\hat a^\dagger_{\lambda,-}(\theta) &= -i \sin(\theta)\hat a^\dagger_{\lambda, \mathrm{H}} \!&+~ \:\cos(\theta)\hat a^\dagger_{\lambda, \mathrm{V}}, \\
\hat a_{\lambda,+}(\theta)         &=~~~ \cos(\theta)\hat a_{\lambda, \mathrm{H}}         \!&+~ i \sin(\theta)\hat a_{\lambda, \mathrm{V}}, \\
\hat a_{\lambda,-}(\theta)         &= -i \sin(\theta)\hat a_{\lambda, \mathrm{H}}         \!&+~ \:\cos(\theta)\hat a_{\lambda, \mathrm{V}}. \\
\end{aligned}
\end{equation}

The symbols $+$ and $-$ denote two orthonormal polarization modes obtained by rotating the standard horizontal and vertical basis vectors. This rotation is defined by an angle $\theta$, which captures the internal structural asymmetry of the \gls{QD} that leads to \gls{FSS}. The resulting polarization modes "$+$" and "$-$" are mutually orthogonal and span the same space as the original $\mathrm{H}$ and $\mathrm{V}$ modes. These rotated modes are introduced here to align the photon creation operators with the natural polarization basis of the \gls{QD} system. Both will be required to model the biexciton-exciton cascade, where each exciton decays into photons with different polarization characteristics.
The full derivation is explained in Appendix \ref{appendix:rotated_ladder_operators}.

\subsection{Total Hilbert Space}\label{subsection:total_Hilbert_space}

To describe the full dynamics of the \gls{QD} interacting with light, we define the total Hilbert space as the tensor product of the \gls{QD} internal states and the relevant photonic modes
\begin{equation}\label{eq:TotalHilbertSpace}
    \mathcal{H}_{\text{total}} = \HQD \bigotimes_{j} \mathcal{H}_{\lambda_j},
\end{equation}
where each $\mathcal{H}_{\lambda_j}$ is the Hilbert space of a spatio-temporal light mode with fixed wavelength $\lambda_j$.

The number of spatio-temporal light modes $\lambda_j$ included depends on the transitions we aim to model. For instance, a biexciton cascade requires at least two distinct spatio-temporal modes to represent the sequential processes, because -- in the absence of \gls{FSS} -- exciton states remain degenerate and a minimal model (compare Figure \ref{fig:XX-X-Cascade}, left) with two spatio-temporal light modes suffices. 

When \gls{FSS} is present, polarization-resolved detection becomes necessary because the exciton eigenstates $\ket{\mathrm{X}_1}$ and $\ket{\mathrm{X}_2}$ couple to orthogonal (rotated) linear polarizations and are split in energy by $\Delta$. This split induces that each transition couples to a separate spatio-temporal mode. Depending on $\Delta$, the spectral modes associated with the two excitons generally overlap when considering a realistic pulse shape. In the limit of large \gls{FSS}, the photons populate nearly orthogonal frequency modes, while for small \gls{FSS} the spectra overlap and the photons are only partially distinguishable in frequency. To capture this behaviour, we represent each emission pulse by two spatio-temporal modes (according to the \gls{FSS}), each of which is resolved into two orthogonal polarizations. This results in four Fock spaces per emission, which allows the model to account for arbitrary spectral overlap of the photon wavepackets. 

In our full mode-resolved description with 4 spatio-temporal spaces, the cascade state remains fully entangled. \gls{FSS} correlates polarization with frequency, so the joint state is entangled across both degrees of freedom. In addition, the energy splitting $\Delta$ causes the exciton state in the \gls{QD} to precess during its lifetime, which imprints a time-dependent relative phase between the two decay paths (see Equation \eqref{eq:time-dependent_photon_release} and Figure \ref{fig:XX-X-Cascade}, right). If this phase evolution is not resolved (e.g. due to detector timing jitter or integration over the exciton lifetime), the oscillatory cross terms in the polarization subspace average out. Entanglement then appears reduced only when spectral and temporal information is ignored, \ie when the frequency modes are traced out. Thus \gls{FSS} does not destroy the underlying photon entanglement, but redistributes it between polarization, frequency, and time.

\subsection{System Hamiltonian}\label{subsec:hamiltonian}

We model the dynamics on $\mathcal{H}_\text{total}$ with a time-dependent Hamiltonian composed of three terms:
\begin{equation}\label{eq:total_hamiltonian}
    \hat {H}(t) = \hat H_\text{FSS} + \hat H_\text{drive}(t) + \hat H_\text{detuning}(t).
\end{equation}
In our model only a classical drive for \gls{TPE} is used to coherently couple to the two-photon transition from $\ketG$ to $\ketXX$ (see Figure \ref{fig:ResonantTwo-PhotonExcitation}). No quantized input fields are included. This is the effective two-photon Hamiltonian that couples $\ketG$ and $\ketXX$ via virtual exciton states, as given for example in
\cite[Eq.3]{ResonantTwoPhotonExcitation3}. This model is valid for large single-photon detuning (the detuning of the laser frequency from the exciton transition), so the exciton manifold is only virtually populated.

In the following, the effective \gls{TPE} Hamiltonian serves as our default model, the \gls{DPE} scheme is included for completeness and comparison.

Throughout this work, the Hamiltonian is written in the rotating frame of the driving field and derived under the rotating-wave approximation, such that rapidly oscillating counter-rotating terms are neglected.

\subsubsection{Exciton Fine-Structure Splitting}

In the bright exciton subspace $\{\ketXone, \ketXtwo\}$ we model the fine-structure splitting by
\begin{equation}\label{eq:H_fss}
    \hat H_\text{FSS} = \frac{\Delta}{2} \Big( \sigmaOp{\mathrm{X}_1}{\mathrm{X}_1}-\sigmaOp{\mathrm{X}_2}{\mathrm{X}_2} \Big).
\end{equation}

Under this Hamiltonian the exciton states acquire phases $\ketXone \to e^{-i\Delta t/2}\ketXone$ and $\ketXtwo \to e^{+i\Delta t/2} \ketXtwo$ during the exciton lifetime. The global phase is irrelevant, so the observable effect is a relative phase factor $e^{i\Delta t/\hbar}$ between the two decay paths. In the biexciton-exciton cascade this phase is inherited by the photon pair and appears as the exponential factor in the second term of Equation \eqref{eq:time-dependent_photon_release}.

\subsubsection{Classical Two-Photon Drive}

We coherently couple $\ketG$ and $\ketXX$ with a semi-classical two-photon Rabi term
\begin{equation}\label{eq:H_flip}
    \hat H_\text{drive}(t)
    = \frac{\hbar\Omega(t)}{2}
    \Big (\sigmaOp{\mathrm{XX}}{\mathrm{G}}+\sigmaOp{\mathrm{G}}{\mathrm{XX}} \Big),
\end{equation}
where $\Omega(t)$ is real and proportional to the square of the classical field envelope, $\Omega(t)\propto |E(t)|^2$. The pulse area $A=\int_{-\infty}^{+\infty}\Omega(t)dt$ controls the $\ketG\leftrightarrow\ketXX$ population transfer.

\subsubsection{Two-Photon Detuning}

The two-photon detuning describes the mismatch between twice the pump frequency and the biexciton transition energy.
Let the instantaneous two-photon detuning be
\begin{equation}\label{eq:Delta2gamma_def}
  \Delta_{2\gamma}=2\omega_{\text{pump}}-\frac{E_{\mathrm{XX}}-E_{\mathrm{G}}}{\hbar},
\end{equation}
where the $\omega_{\text{pump}}$ is the angular frequency of the pump laser.

In our effective model this detuning is treated as a time-independent energy shift of the biexciton level,
\begin{equation}\label{eq:H_detuning}
  \hat H_\text{detuning}
  = \hbar\frac{\Delta_{2\gamma}}{2}\sigmaOp{\mathrm{XX}}{\mathrm{XX}}.
\end{equation}
Equivalently, one may add $-\hbar\,\tfrac{\Delta_{2\gamma}(t)}{2}\op{\mathrm{G}}{\mathrm{G}}$ instead; both forms differ by a scalar and are physically identical.

This treatment does not describe non-resonant \gls{TPE} with intermediate nonradiative decay via phonons (see Figure \ref{fig:Phonon-AssistedTwo-PhotonExcitation}), but simply accounts for the continuous dependence of the pump-biexciton coupling on the laser wavelength. In this way, we not only fix the detuning across the entire laser pulse -- \ie the pulse is assumed to have a well-defined frequency profile throughout its duration -- but also model that slightly detuned pulses with realistic spectral shapes can couple to to a transition with a certain probability.

\subsubsection{Dichromatic Pulsed Excitation Hamiltonian}

Beyond the effective \gls{TPE} scheme, the biexciton may also be populated via a two-colour excitation scheme as defined in Section \ref{section:Physics}.\ref{sec:phys-dichromatic-excitation}.
In this case, the system Hamiltonian acquires additional drive terms \cite[Eq.1]{BiexcitonInQDviaARP} of the form
\begin{equation}
\hat H_\text{drive, DPE}(t)
=
\hat H_{\mathrm{G}\leftrightarrow\mathrm{X}}(t)
+
\hat H_{\mathrm{X}\leftrightarrow\mathrm{XX}}(t),
\end{equation}
\begin{align}
\hat H_{\mathrm{G}\leftrightarrow\mathrm{X}}(t)
&=
\frac{\hbar}{2}\sum_{j \in \{1,2\}}
\Big[
\Omega_{\mathrm{G},\mathrm{X}_j}(t)\,\sigmaOp{\mathrm{G}}{\mathrm{X}_j}
+\text{h.c.}
\Big],\\
\hat H_{\mathrm{X}\leftrightarrow\mathrm{XX}}(t)
&=
\frac{\hbar}{2}\sum_{j \in \{1,2\}}
\Big[
\Omega_{\mathrm{X}_j,\mathrm{XX}}(t)\,\sigmaOp{\mathrm{XX}}{\mathrm{X}_j}
+\text{h.c.}
\Big].
\end{align}

Depending on the choice of pulse shapes, relative timing, and detunings, this description encompasses resonant \gls{TPE} excitation as well as \gls{ARP} chirped pulse excitation protocols.

In this formulation, single-photon and two-photon detunings associated with the individual transitions may be absorbed into the rotating frame or incorporated into the complex phases of the time-dependent Rabi frequencies $\Omega_{\mathrm{G},\mathrm{X}_j}(t)$ and $\Omega_{\mathrm{X}_j,\mathrm{XX}}(t)$. The explicit two-photon detuning $\Delta_{2\gamma}$ is specific to the effective \gls{TPE} model and is not required when treating \gls{DPE} or \gls{ARP} excitation schemes.

\subsubsection{Identification of the System Hamiltonian}\label{subsec:system_hamiltonian}

For the treatment of the phonon-induced decoherence, further analysed in Section \ref{section:Mathematics}.\ref{subsec:phonon-induced_dissipation}, we follow the polaron-frame master equation formalism. In this formalism the total Hamiltonian is decomposed into a system part, a phonon bath, and a residual system-bath interaction.
In this framework, the system Hamiltonian $\hat{H}_S(t)$ is identified with the \gls{QD} Hamiltonian Eq.~\eqref{eq:total_hamiltonian}, restricted to the electronic degrees of freedom of the \gls{QD} and expressed in the rotating frame.

After the polaron transformation, $\hat{H}_S(t)$ retains the same operator structure, while coherent couplings are renormalized according to:
\begin{equation}
\Omega(t)\rightarrow\Omega_r(t)=\Omega(t) \expval{B}\!(T),
\end{equation}
with thermal dressing factor $\expval{B}\!(T)$.

Phonon-induced fluctuations then give rise to additional dissipative terms in the Lindblad master equation, whose form is determined by the instantaneous eigenstructure of $\hat{H}_S(t)$.
This identification is standard in the description of driven systems with \gls{QD}-phonon coupling and applies generally to pulsed, chirped, and multi-color excitation schemes \cite{mccutcheonQuantumDotRabi2010, reiterRolePhononsExciton2014, ramsayDampingExcitonRabi2010}.

\subsection{Lindblad Master Equation}\label{sec:lindblad}

To describe the full dynamics of the \gls{QD} including spontaneous emission and decoherence effects, we transition from the unitary evolution governed by the Schrödinger equation to a density matrix formalism, using the Lindblad master equation:
\begin{equation}
     \frac{d\rho}{dt} = -\frac{i}{\hbar}[\hat{H}(t), \rho] + \sum_j \mathcal{D}[\hat{L}_j]\rho,
\end{equation}

where $\hat{H}(t)$ is the total Hamiltonian, including excitation and drive terms, $\rho$ is the density operator of the system that describes the state. The term $\mathcal{D}[\hat{L}_j]$ is the Lindblad superoperator:
\begin{equation}\label{eq:lindblad}
\mathcal{D}[\hat{L}_j](\rho) = \hat{L}_j\rho \hat{L}_j^\dagger - \frac{1}{2} \{\hat{L}_j^\dagger \hat{L}_j, \rho\},
\end{equation}
and $\hat{L}_j$ are jump operators representing irreversible -- \ie non-unitary -- processes (\eg spontaneous emission). 
\cite{TheoryOfOpenQuantumSystems}

\subsubsection{Radiative Dissipation}

Each spontaneous emission event from the \gls{QD} is modelled by a collapse (jump) operator \(\hat{L}_j\), representing a transition from an excited state to a lower state, accompanied by the emission of a photon into a specific photonic mode. The general form is:
\begin{equation}
\hat{L}_k = \sqrt{\gamma_k}\, \hat{a}^\dagger_{\lambda_k}(\theta_k) \otimes \sigmaOp{f_k}{i_k}.
\end{equation}

Here, $\gamma_k$ denotes the spontaneous emission rate associated with the transition $\ket{f_k} \to \ket{i_k}$ within the \gls{QD}. The operator $\sigmaOp{f_k}{i_k}$ describes this transition, while $\hat{a}^\dagger_{\lambda_k}(\theta_k)$ represents the photon creation operator in the rotated polarization basis defined by angle $\theta_k$.

For the biexciton cascade system, the collapse operators become:
\begin{align} \label{eq:collapse}
    \begin{split}
    \hat{L}_{\mathrm{XX}\to \mathrm{X}} &= \sqrt{\gamma_{\mathrm{XX}\to \mathrm{X}_1}}\sigmaOp{\mathrm{XX}}{\mathrm{X}_1}\otimes\hat{a}_{\lambda_1, -}^\dagger\left(\theta\right) \\
    &+ \sqrt{\gamma_{\mathrm{XX}\to\mathrm{X}_2}}\sigmaOp{\mathrm{XX}}{\mathrm{X}_2}\otimes \hat{a}_{\lambda_2, +}^\dagger\left(\theta\right),
    \end{split}\\
    \begin{split}
    \hat{L}_{\mathrm{X}\to \mathrm{G}}  &= \sqrt{\gamma_{\mathrm{X}_1\to \mathrm{G}}}\sigmaOp{\mathrm{X}_1}{\mathrm{G}}\otimes\hat{a}_{\lambda_3,-}^\dagger\left(\theta\right) \\
    &+ \sqrt{\gamma_{\mathrm{X}_2\to\mathrm{G}}}\sigmaOp{\mathrm{X}_2}{\mathrm{G}}\otimes \hat{a}_{\lambda_4,+}^\dagger\left(\theta\right).
    \end{split}
\end{align}

\paragraph{Note on Mode Assignments and Fine Structure Splitting}

When the \gls{FSS} vanishes ($\Delta = 0$), the two exciton levels become degenerate, and their associated transitions may emit photons into the \textit{same} photonic mode and polarization state, for example $\lambda_1 = \lambda_2$ and $\hat{a}^\dagger_{\lambda_1,+} = \hat{a}^\dagger_{\lambda_2,+}$. In such cases, interference effects can arise, and the emitted photons may be partially indistinguishable. These interference effects are the coherent addition of indistinguishable decay amplitudes from the two cascade paths $\mathrm{XX}\!\to\!\mathrm{X}_1\!\to\!\mathrm{G}$ and $\mathrm{XX}\!\to\!\mathrm{X}_2\!\to\!\mathrm{G}$. When $\Delta=0$ and the two transitions emit into the same optical mode (\ie same wavelength and polarization), the corresponding jump operator has the form 
\begin{equation}
    \hat L_{\mathrm{X}\to\mathrm{G}} \propto \hat a^\dagger \!\otimes\! \big(\sigmaOp{\mathrm{X}_1}{\mathrm{G}} + \sigmaOp{\mathrm{X}_2}{\mathrm{G}}\big),
\end{equation}
so cross terms such as $\sigmaOp{\mathrm{X}_1}{\mathrm{G}}\,\rho\,\sigmaOp{\mathrm{G}}{\mathrm{X}_2}$ survive in the Lindbladian. These coherences generate the observed polarization interference and yield a coherent two-photon state (\eg $\ket{\mathrm{HH}}+e^{i\phi}\ket{\mathrm{VV}}$ in a suitable basis). Two mechanisms reduce these cross terms: 

(i) \emph{genuine which-path information}, \eg spectral distinguishability due to \gls{FSS} (or orthogonal spatial/polarization modes), which makes the two decay channels in principle distinguishable,

(ii) \emph{phase averaging}, where the \gls{FSS}-induced relative phase $e^{i\Delta t/\hbar}$ is not time/frequency resolved, so averaging over the exciton dwell time suppresses the off-diagonals and lowers the observed entanglement visibility. 

In contrast, for $\Delta \neq 0$, the transitions produce photons of distinguishable frequency and the modes $\lambda_k$ must be treated as distinct. This distinction is crucial for correctly modelling entanglement and photon interference behaviour in the system.

\subsubsection{Phonon-Induced Dissipation}\label{subsec:phonon-induced_dissipation}

Phonon interactions lead to temperature-dependent decoherence and scattering processes 
that are not directly observable in the photonic environment and are therefore modelled as effective dissipative channels acting on the \gls{QD}'s degrees of freedom.

To accurately capture phonon effects in the presence of coherent driving and super-Ohmic acoustic phonon environments, we employ a polaron-frame master equation approach.~\cite{reiterRolePhononsExciton2014, mccutcheonQuantumDotRabi2010, ramsayDampingExcitonRabi2010}

In this approach, the diagonal exciton-phonon coupling is treated non-perturbatively via a unitary polaron transformation
\begin{equation}\label{eq:polaron_transformation}
    U_\text{pol}=\exp\!\left[\sum_{\nu}
    \sigmaOp{\nu}{\nu}\sum_q\frac{g_q^\nu}{\omega_q}\left(\hat b_q^ \dagger - \hat b_q\right)\right],
\end{equation}
while residual phonon-induced fluctuations are handled within a Born-Markov approximation. In Eq.~\eqref{eq:polaron_transformation} $\nu\in\{\mathrm{X}_1, \mathrm{X}_2, \mathrm{XX}\}$ and ladder operators $\hat{b}^\dagger_q$, $\hat{b}_q$ operate on the phononic modes labelled by $q$. Additionally, $\omega_q$ denotes the angular frequency of the phonon mode with a wavevector $q$, and $g_q^\nu$ is the exciton-phonon coupling strength between the phonon mode $q$ and quantum state $\ket{\nu}$.

The resulting reduced dynamics remains time-local and can be written in \gls{GKSL} form. Phonon effects therefore enter the Lindblad master equation through additional effective collapse operators with temperature-dependent rates, while radiative decay channels remain unchanged.

\paragraph{Temperature Dependence}

The temperature dependence is governed by the polaron phase function \cite[Eq.20]{mccutcheonQuantumDotRabi2010}
\begin{equation}
\begin{aligned}
\phi(t,T)=\int_0^\infty \frac{J(\omega)}{\omega^2}
\Bigg[
&\coth\!\left(\frac{\hbar\omega}{2k_{\mathrm B}T}\right)\cos(\omega t) \\
&-\, i\sin(\omega t)
\Bigg] d\omega,
\end{aligned}
\end{equation}
where $J(\omega)$ denotes the phonon spectral density. Following Ref.~\cite{reiterRolePhononsExciton2014}, we may model the phonon spectral density in a super-Ohmic form as $J(\omega)=\alpha\,\omega^3\exp[-(\omega/\omega_c)^ 2]$. In a harmonic-confinement model one has $\omega_c=\sqrt{2}\,c_s/a$ and $\alpha\propto|D_e-D_h|^2/(\rho\,\hbar\,c_s^5)$. For self-assembled \ch{In(Ga)As}/\ch{GaAs} \glspl{QD}, typical parameters used in a polaron-frame modelling are $\alpha\approx\qty{0.027}{\pico\second\squared}$ and $\omega_c\approx\qty{2.2}{\pico\second^{-1}}$ \cite[Sec.3.2]{mccutcheonQuantumDotRabi2010}.

Then, the thermal dressing factor can be computed via:
\begin{equation}
\expval{B}\!(T)=
\exp\left[
-\frac{1}{2} \int_0^\infty \frac{J(\omega)}{\omega^2}
\coth\left(\frac{\hbar\omega}{2k_{\text{B}}T}\right) d\omega
\right],
\end{equation}
and it renormalizes coherent couplings in the polaron frame.

The temperature dependence is introduced into the coherent drive Hamiltonian $\hat{H}_\text{drive}(t)$ from Eq.~\eqref{eq:H_flip}, by substituting the term $\Omega(t)$ by the renormalized $\Omega_r(t)$, with
\begin{equation}
    \Omega_r(t) = \Omega(t)\expval{B}\!(T).
\end{equation}
More generally, each driven transition $i\leftrightarrow j$ is renormalized by $\expval{B_{ij}}$, we write $\expval{B}$ for brevity.

\paragraph{Phonon-Induced Scattering Rates}

In addition to the coherent renormalization $\Omega(t)\mapsto\Omega_r(t)=\Omega(t)\expval{B}$, residual phonon effects are incorporated by adding time-local \gls{GKSL} dissipators whose operator structure is fixed by the driven transition, while the corresponding rates depend on temperature and on the instantaneous drive and detuning. Concretely, for each driven transition $ i \leftrightarrow j$ we define the dressed splitting
\begin{equation}
    \omega^*(t)=\sqrt{\Delta(t)^2+\Omega_R(t)^2},
\end{equation}
where $\Delta(t)$ is the instantaneous detuning and $\Omega_R(t)=\lvert\Omega_r(t)\rvert$ is the magnitude of the polaron-renormalized Rabi frequency.

Phonon-induced scattering is represented by three collapse operators,
\begin{equation}
\begin{aligned}
    L_{j\leftarrow i}(t) &= \sqrt{\gamma_{j\leftarrow i}(t)}\,\sigmaOp{j}{i}, \\
    L_{i\leftarrow j}(t) &= \sqrt{\gamma_{i\leftarrow j}(t)}\,\sigmaOp{i}{j}, \\
    L_{\text{cd}}^{(ij)}(t) &= \sqrt{\gamma_{\text{cd}}^{(ij)}(t)}\left(
    \sigmaOp{j}{j}-\sigmaOp{i}{i}\right).
\end{aligned}
\end{equation}
The first two terms describe phonon-assisted population transfer between the states $i$ and $j$, while the third term accounts for phonon-induced cross-dephasing. The associated rates $\gamma_{j\leftarrow i}(t)$, $\gamma_{i\leftarrow j}(t)$, and $\gamma_{\text{cd}}$ are obtained from integrals over the polaron bath correlation function
\begin{equation}
    G(\tau,T)=e^{\phi(\tau,T)}-1,
\end{equation}
evaluated at the instantaneous dressed splitting $\omega^*(t)$. Explicitly, the rates are proportional to real parts of correlation integrals of the form
\begin{equation}
    \Re\!\int_0^\infty e^{\pm i\omega^*(t)\tau}G(\tau,T)\,d\tau,
\end{equation}
while the cross-dephasing channel involves a real correlation integral constructed from $1-\exp[-\Re\phi(\tau,T)]$.

This construction follows standard polaron master-equation treatments and captures the dominant temperature- and drive-dependent phonon-assisted scattering and decoherence mechanisms relevant for the biexciton-exciton cascade, while keeping the reduced dynamics explicitly time-local and in \gls{GKSL} form.

Non-Markovian phonon memory effects beyond this time-local generator are neglected.

\paragraph{Exciton Relaxation}

In addition to drive-induced phonon scattering, we optionally include phonon-assisted population transfer between the exciton eigenstates $\mathrm{X}_1$ and $\mathrm{X}_2$. Writing the exciton splitting as $\Delta E=\sqrt{\Delta^2+(2\delta')^2}$ with \gls{FSS} $\Delta$ and off-diagonal bright-exciton mixing term $\delta'$, we define $\omega_{\mathrm{X}}=\Delta E/\hbar$. 
The corresponding rates are modelled using a minimal golden-rule estimate (see, \eg polaron master-equation treatments) \cite{quilterPhononassistedPopulationInversion2014, royInfluenceElectronacousticPhonon2011},
\begin{align}
\begin{split}
\gamma_{\downarrow} &=2\pi s^2 J(\omega_{\mathrm{X}}) \big( n(\omega_{\mathrm{X}}) + 1 \big),\\
\gamma_{\uparrow}   &=2\pi s^2 J(\omega_{\mathrm{X}}) n(\omega_{\mathrm{X}}),
\end{split}
\end{align}
with $s^2=(\phi_{\mathrm{X}_1}-\phi_{\mathrm{X}_2})^2$ and Bose factor $n(\omega)=\big(e^{\hbar\omega/(k_{\mathrm{B}} T)}-1\big)^{-1}$. 
The uphill and downhill rates introduced here are corresponding to the state-resolved scattering rates $\gamma_{j\leftarrow i}$ when the exciton's energy ordering is fixed.
These rates enter the master equation via jump operators $\sqrt{\gamma}\,\op{\mathrm{X}_j}{\mathrm{X}_i}$. 
Here, $\gamma_\downarrow$ ($\gamma_\uparrow$) denotes phonon-assisted relaxation from the higher- (lower-) energy exciton eigenstate to the lower- (higher-) energy one. These rates are not independent objects, but correspond directly to the state-resolved scattering rates $\gamma_{j\leftarrow i}$ introduced above. In the present model, the exciton eigenstates are ordered such that $\mathrm{X}_1$ is the higher-energy state and $\mathrm{X}_2$ the lower-energy state, so that $\gamma_\downarrow \equiv \gamma_{\mathrm{X}_2\leftarrow\mathrm{X}_1}$ and $\gamma_\uparrow \equiv \gamma_{\mathrm{X}_1\leftarrow\mathrm{X}_2}$.
\section{Simulation}\label{section:Simulation}

This section describes how \gls{TPE} in the large single-photon detuning regime is implemented in practice. The simulation propagates the joint state of the \gls{QD} and its photonic environment $\mathcal{H}_\text{total}$ under the time-dependent Hamiltonian of Section~\ref{section:Mathematics}.\ref{subsec:hamiltonian} together with radiative Lindblad jumps. The infinite-dimensional Fock spaces $\mathcal{F}_{\lambda, p}$ are truncated by the user to a cutoff $N_\text{cut}$ (typically $N_\text{cut}=2$), which restricts each mode to the subspace $\{\ket{0},\ket{1}\}$ and is sufficient for the cascade dynamics considered here.

The system is initialized in the ground state of the \gls{QD} with vacuum in all photonic modes,
\begin{equation}
    \rho(t_0) = \op{\mathrm{G}}{\mathrm{G}} \otimes \op{\text{vac}}{\text{vac}}.
\end{equation}

The solver integrates from $t_0=0$ to a final time $t_\text{f}$ chosen by the user ideally so that the entire excitation pulse and all cascade emissions are contained in $[t_0, t_\text{f}]$. At $t_\text{f}$, the \gls{QD} state is traced out to yield the reduced photonic state.
\begin{equation}
    \rho_{\gamma}(t_\text{f})= \mathrm{Tr}_\mathrm{QD}\rho(t_\text{f}).
\end{equation}
This state is then processed by the software module to extract the \gls{CPTP} quantum channel in the form of a Kraus map, which maps the trivial input space to the evolved state of the output photonic modes at the chosen final time.

\subsection{Drive Definition}\label{subsec:drive}

The \gls{TPE} is implemented using a classical optical field drive. Rather than specifying a Rabi frequency directly, the drive is defined at the level of a physical electric field, allowing amplitude, polarization, and frequency chirp to be encoded explicitly.

The drive is specified by a dimensionless temporal envelope $f(t)$, a peak electric-field amplitude $E_0 [\si{\volt\per\meter}]$, and an optional optical carrier. The physical field envelope is
\begin{equation}
    E_{\text{env}}(t) = E_0 f(t),
\end{equation}
with physical time $t$. The solver evolves in rescaled time $t' = t/s$, where the factor $s [\si{\second}]$ is set by the \texttt{time\_unit\_s} parameter; all quantities are internally mapped to solver units.

The optical carrier defines the instantaneous angular frequency
\begin{equation}
    \omega_{\text{L}}(t) = \omega_0 + \delta\omega(t),
\end{equation}
where $\omega_0$ is a constant base frequency and $\delta\omega(t)$ may be a constant offset or a time-dependent chirp.

The polarization of the driving field is described by a Jones vector in the fixed $\{\mathrm{H},\mathrm{V}\}$ basis, optionally followed by a Jones matrix that represents linear polarization optics (\eg rotations or wave plates). The resulting effective polarization determines how the field couples to the optical transitions of the \gls{QD}.

The drive object itself is \gls{QD}-agnostic: it provides the instantaneous field amplitude, frequency, and polarization, while the construction of the two-photon coupling between $\ketG$ and $\ketXX$, including the effective two-photon detuning, is handled internally when assembling the Hamiltonian.

\subsection{Quantum Dot Parametrization and Modelled Physical Effects}\label{subsec:qd_params}

While Section~\ref{section:Mathematics} specifies the dynamical model, the implementation requires a concrete parametrization of the \gls{QD} and its environment. In practice, the simulator constructs a four-level \gls{QD} with states $\{\ketG, \ketXone, \ketXtwo, \ketXX\}$ and consumes the following parameter blocks.

\subsubsection{Energy Structure}

The electronic energies are specified by an \texttt{EnergyStructure} object, parametrized by the exciton center energy $E_{\mathrm{X}}$, \gls{FSS} $\Delta$, and biexciton binding energy $E_\text{b}$. Internally, the levels are instantiated as
\begin{align}
\begin{split}
E_{\mathrm{X}_1} &= E_{\mathrm{X}} +\frac{\Delta}{2},\\
E_{\mathrm{X}_2} &= E_{\mathrm{X}} - \frac{\Delta}{2},\\
E_{\mathrm{XX}}  &= 2E_{\mathrm{X}} - E_{\text{b}},
\end{split}
\end{align}
with $E_\mathrm{G} \equiv 0$. These energies determine the transition frequencies used in the Hamiltonian and in the collapse operators.

\subsubsection{Optical Dipoles and Polarization Selection Rules}

Radiative couplings are specified by \texttt{DipoleParams}. The model associates each allowed transition with (i) a dipole magnitude $\mu$ and (ii) a Jones polarization vector $\mathbf{e}_{\text{pol}}$ in the fixed $\{\mathrm{H},\mathrm{V}\}$ basis. Conceptually, each directed transition carries a dipole vector
\begin{equation}
\mathbf{d}_{S_1\rightarrow S_2} = \mu_{S_1S_2} \mathbf{e}_{\text{pol},S_1S_2}.
\end{equation}
For $\mathrm{XX}$-$\mathrm{X}$-cascade simulations we provide a convenience constructor that generates the polarization selection rules from the \gls{FSS} and an optional exciton-mixing parameter $\delta'$. In the absence of mixing, the exciton eigenstates align with the $\mathrm{H}$/$\mathrm{V}$ axes; finite $\delta'$ rotates the eigenbasis and yields elliptical polarization eigenmodes. This parametrization determines how \gls{FSS} redistributes emission into polarization-resolved photonic modes.

\subsubsection{Cavity-Enhanced Spontaneous Emission}

The radiative rates entering the Lindblad operators are set through \texttt{CavityParams}, defined by the cavity quality factor $Q$, effective mode volume $V_{\text{eff}}$, cavity wavelength $\lambda_{\text{cav}}$, and refractive index $n$. These parameters determine the Purcell enhancement factor \cite{cuiQuantumEfficiencySinglephoton2005}
\begin{equation}
F_{\text{p}}=\frac{3}{4\pi^2}\left(\frac{\lambda_{\text{cav}}}{n}\right)^3\frac{Q}{V_{\text{eff}}},
\end{equation}
which scales the emission rates $\gamma_{S_1\to S_2}$ used by the \texttt{CollapseBuilder}. In the present work we use this cavity model to set physically consistent decay rates without introducing additional fitting parameters.

\subsubsection{Phonon Environment (Optional)}

The simulator can optionally include a lattice bath via \texttt{PhononParams}. The current implementation supports a deformation-potential \gls{LA} polaron model with spectral density
\begin{equation}
J(\omega) = \alpha\omega^3\exp\left[-\left(\frac{\omega}{\omega_{\text{c}}}\right)^2\right],
\end{equation}
specified by the strength $\alpha$ and cutoff frequency $\omega_\text{c}$, together with the lattice temperature $T$. Effective displacement parameters $\phi_\mathrm{G},\phi_{\mathrm{X}_1}$, $\phi_{\mathrm{X}_2}$, and $\phi_{\mathrm{XX}}$ control the coupling strengths through differences $(\phi_i-\phi_j)^2$. Depending on configuration, the phonon model can provide (i) polaron renormalization factors $\expval{B}\!(T)$ for optical couplings, (ii) phenomenological constant dephasing/relaxation rates, and (iii) parameters for drive-induced dephasing and scattering terms that are evaluated in a downstream stage from the instantaneous drive amplitude.

\subsubsection{Summary}

Taken together, the parameter blocks \texttt{EnergyStructure}, \texttt{DipoleParams}, \texttt{CavityParams}, and (optionally) \texttt{PhononParams} fully specify the physical \gls{QD} and its environment. From these inputs the implementation constructs the Hamiltonian coefficients, radiative collapse operators, and any enabled phonon contributions, and then propagates the composite state as described above.

While this flexibility enables the exploration of a wide range of parameter regimes and excitation protocols, the validity of the resulting dynamics ultimately depends on the physical consistency of the chosen model parameters. The simulation framework does not enforce regime constraints, and care must therefore be taken to ensure that the selected parameters remain within the domain of applicability of the underlying approximations.

\subsection{Hamiltonian and Collapse Operators}

All single-system operators and composite operators are instantiated using
\texttt{PhotonWeave} \cite{PhotonWeave}, which also constructs the composite
Hilbert space $\mathcal{H}_\text{total}$ from the \gls{QD} states and the
implicit spatio-temporal photonic modes (two or four, depending on the presence
of \gls{FSS}).

The Hamiltonian of Section~\ref{section:Mathematics}.\ref{subsec:hamiltonian} is assembled by a staged construction pipeline. First, a drive decoder maps the user-defined classical field (envelope, amplitude, carrier, polarization) onto the relevant optical transitions of the \gls{QD}. This determines which flip operators appear (\eg $\sigmaOp{\mathrm{G}}{\mathrm{XX}}$ for \gls{TPE}) and fixes the effective detuning relative to the corresponding transition energies.

Second, the decoder evaluates the effective coupling strengths on the solver time grid, including renormalization factors if a phonon model is enabled. The resulting coefficients are then used to assemble the time-dependent drive Hamiltonian terms, while static energy shifts (such as the two-photon biexciton detuning of the excitation pulse) are added as time-independent operators.

The final Hamiltonian is composed of the following contributions:
\begin{itemize}[noitemsep, topsep=0pt]
    \item \textbf{Static Hamiltonian Terms:} The \gls{FSS} term
    $\hat H_\mathrm{FSS}$ (see Eq.~\eqref{eq:H_fss}) and static detuning shifts such as
    the biexciton energy offset (see Eq.~\eqref{eq:H_detuning}).

    \item \textbf{Drive Hamiltonian Terms:} Time-dependent flip operators
    (see Eq.~\eqref{eq:H_flip}) with coefficients derived from the decoded classical
    field and evaluated in solver time $t' = t/s$.

    \item \textbf{Phonon-Induced Collapse Terms (Static):} If enabled, the phonon
    model contributes time-independent Lindblad operators describing incoherent
    dephasing and thermal relaxation between electronic states (\eg relaxation
    within the exciton manifold). These terms represent drive-independent
    coupling to the lattice and are active throughout the evolution.

    \item \textbf{Phonon-Induced Drive-Dependent Terms (Optional):} Additional
    dephasing or scattering processes whose rates depend on the instantaneous
    drive strength are constructed as time-dependent contributions when the
    corresponding model options are enabled.
\end{itemize}

Spontaneous emission is described by Lindblad jump operators of the form in Eq.~\eqref{eq:collapse}. These are assembled by a \texttt{CollapseBuilder}, which couples each \gls{QD} transition $\sigmaOp{S_1}{S_2}$ to the corresponding photon creation operator $\hat a^\dagger_{\lambda,\pm}(\theta)$ with rate $\gamma_{S_1\to S_2}$ determined from the dipole and cavity parameters (\texttt{DipoleParams},\texttt{CavityParams}). The complete set of Hamiltonian and collapse operators is passed directly to the master-equation solver.

\subsection{Quantum Channel}\label{subsec:channel}

At the final time $t_{\text{f}}$ we trace out the \gls{QD} to obtain the reduced photonic state $\rho_{\gamma}(t_{\text{f}}) \equiv \rho_{\gamma}$. Because the input space is trivial (a fixed pure input: $\ket{G}\otimes\ket{\text{vac}}$), the overall evolution implements a \emph{state-preparation channel} from a one-dimensional input space $\mathbb{C}$ to the photonic output space.
Any such channel admits a Kraus representation
\begin{equation}
        \Phi(\cdot) = \sum_j \hat{K}_j(\,\cdot\,) \hat{K}_j^\dagger, \quad \sum_j \hat{K}_j^\dagger \hat{K}_j = \hat {I}_\mathbb{C}=1.
\end{equation}

Since the input is one-dimensional, each $K_j$ is a column vector on the photonic Hilbert space, and the channel acting on the scalar "state" $\op{0}{0}$ reduces to
\begin{equation}\label{eq:channel_from_rho}
    \rho_{\gamma} = \Phi(\op{0}{0}) =
    \sum_j \hat{K}_j \op{0}{0} \hat{K}_j^\dagger=\sum_j \hat{K}_j \hat{K}_j^\dagger.
\end{equation}

A convenient (minimal) Kraus set is obtained from the spectral decomposition of the final state:
\begin{equation}
    \rho_{\gamma} = \sum_{k=1}^r \lambda_k \op{\phi_k}{\phi_k} \qquad \lambda_k \ge0, \sum_k \lambda_k = 1,
\end{equation}
where $r=\mathrm{rank}(\rho_{\gamma})$. 
Define
\begin{equation}\label{eq:kraus_from_eigendecomp}
    \hat{K}_k = \sqrt{\lambda_k}\op{\phi_k}{0},
\end{equation}
then Equation \eqref{eq:channel_from_rho} is satisfied and the trace-preserving condition holds automatically.
This Kraus set is unique up to unitary rotations among the $r$ non-zero eigenmodes. In practice, we compute $\{\lambda_k,\ket{\phi_k}\}$ numerically (and clamp tiny negative eigenvalues to zero within a set tolerance), then store the $\hat{K}_k$ as the exported channel.

\subsection{Metrics} 

The simulation evaluates several figures of merit that quantify both the source brightness and the quality of the emitted entangled state.

\subsubsection{Photon Counting}

The brightness is quantified by the average photon number in the two emissions. For a single mode with annihilation operator $\hat a_{\lambda, p}$ the observable is
\begin{equation}
    N_{\lambda,p} = \mathrm{Tr}\left(
    \hat a_{\lambda,p}^\dagger \hat a_{\lambda, p}\,\rho_{\gamma}
    \right).
\end{equation}

In the absence of \gls{FSS}, the cascade is described by two spatio-temporal modes ("early", "late"). With \gls{FSS} present, the emission is distributed over four Fock spaces corresponding to the rotated polarization basis. In both cases the sum of the average photon numbers recovers the total pair-emission probability.

The absolute brightness is therefore determined only by the excitation conditions, in particular by how closely the driving pulse realizes a $\pi$-pulse on the biexciton transition. \gls{FSS} does not reduce the total number of emitted photons, but redistributes them among polarization- and frequency-resolved modes. Deviations of $N_\text{early}$ and $N_\text{late}$ from unity thus reflect limitations of the excitation:
an insufficient pulse area (less than a $\pi$-pulse) leaves residual vacuum, whereas an excessive or overly long pulse (greater than a $\pi$-pulse or extending beyond the decay times) can generate multi-photon components even beyond the truncation $N_\text{cut}$. Further discrepancies may also arise from numerical issues, for example when the integration window does not fully cover the excitation pulse and decay dynamics, or when the solver time steps are too coarse to resolve its temporal structure.

\subsubsection{Logarithmic Negativity}

Entanglement between the two emitted photons is quantified by the logarithmic negativity, defined as
\begin{equation}
    E_\mathcal{N}(\rho) = \log_2 \| \rho^{T_\text{late}} \|_1,
\end{equation}
where $\rho^{T_\text{late}}$ denotes the partial transpose with respect to the late photon's subsystem and $\|\cdot\|_1$ is the trace norm. In the simulation the reduced photonic state $\rho_{\gamma}$ is bipartitioned into an "early" and a "late" photon (summing over the polarizations), and then the partial transpose is taken with respect to the latter. Two versions are useful in practice.

\emph{Unconditional log-negativity} is evaluated on the full $\rho_{\gamma}$, including vacuum and single-photon components. The quantity reflects end-to-end entanglement available from the source, and is reduced whenever the excitation is imperfect or when numerical integration fails to capture the full pulse.

\emph{Conditional log-negativity} (post-selected) is evaluated after projecting onto the two-photon subspace and renormalizing. In this case the cascade ideally produces a Bell state with ${E_\mathcal{N}}^\text{cond}\approx 1$.

\gls{FSS} by itself does not reduce either measure, since it corresponds only to local unitary rotations between polarization and frequency modes.

\subsubsection{Purity}

The mixedness of the photonic state is characterized by the purity,
\begin{equation}
P(\rho) = \mathrm{Tr}(\rho^2).
\end{equation}

A pure state yields $P=1$, while mixed state gives $P<1$. In the simulation the purity is computed from the reduced photonic density matrix $\rho_{\gamma}$.

\subsubsection{Indistinguishability}

We quantify indistinguishability by the overlap of the single-photon wave packets emitted on the two polarization branches split by the \gls{FSS}. For transform-limited Lorentzian wave packets (corresponding to exponential temporal decay)
with effective coherence decay rate $\gamma_\text{eff}$, the corresponding linewidth in energy units is $\Gamma = \hbar \gamma_\text{eff} [\si{\electronvolt}]$, and with \gls{FSS} $\Delta [\si{\electronvolt}]$, the analytic overlap is
\begin{equation}
  \Lambda = \frac{\Gamma}{\sqrt{\Gamma+\Delta^2}}.
\end{equation}

We evaluate $\Lambda$ separately for the early and late photons using $\Gamma=\tfrac{\hbar}{2}(\gamma_1+\gamma_2)$ from the corresponding radiative rates, and also report their arithmetic mean. The corresponding Hong-Ou-Mandel visibility for two identical sources is $V_\text{HOM}=\Lambda^2$.

\subsection{Implementation Notes}

The simulation is implemented using a combination of open-source Python packages:
\begin{itemize}[noitemsep, topsep=0pt]
    \item \textbf{PhotonWeave} \cite{PhotonWeave}: used to instantiate the Hilbert spaces of the \gls{QD} and the photonic modes, and to construct the operators entering the Hamiltonian and collapse terms.
    \item \textbf{QuTiP} \cite{qutip1, qutip2}: provides the \texttt{mesolve} master-equation solver, which propagates the density matrix under the specified Hamiltonian and Lindblad operators.
    \item \textbf{QSI}: a module provided by the journal special issue that standardizes communication between independent simulation components. Our implementation uses QSI to wrap the simulation as an interoperable form of Kraus map, see Equation \eqref{eq:kraus_from_eigendecomp}. This allows direct integration with other modules.
\end{itemize}

In addition to the core channel interface, the implementation provides diagnostic tools to monitor and validate the simulation. Users can generate population traces of the \gls{QD} states and the photonic modes for any chosen excitation pulse. These diagnostics make it straightforward to verify $\pi$-pulse operation, inspect the cascade dynamics, and identify deviations due to detuning or \gls{FSS}. The plotting utilities are available both in standalone runs and when the module is executed through QSI, ensuring that the simulation remains transparent even in automated workflows.

The source code is openly available at \url{https://github.com/tqsd/BEC} under an open-source license \cite{Dataset}.

\onecolumn

\subsection{Results}

\begin{figure}[H] 
    \centering
    \includegraphics[width=\textwidth]{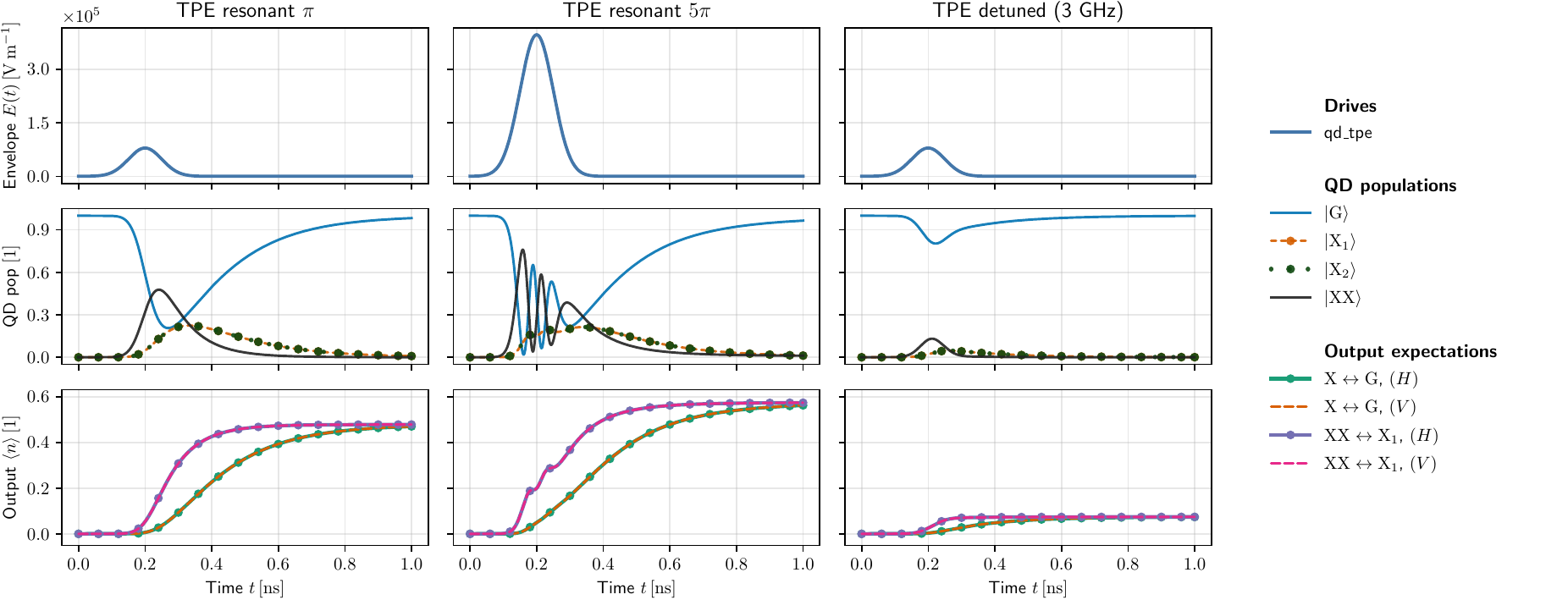}
    \caption{Different exemplary \gls{TPE} scenarios. From left to right: $\pi$-pulse equivalent in resonance with the $\mathrm{XX}$ transition, $5\pi$-pulse equivalent in resonance with the $\mathrm{XX}$ transition, $\pi$-pulse equivalent detuned by \qty{3}{\giga\hertz} from the $\mathrm{XX}$ transition. Phonon dissipation mechanisms were turned off for these simulations.}
    \label{fig:plot_tpe}
\end{figure}

\begin{figure}[H] 
    \centering
    \includegraphics[width=\textwidth]{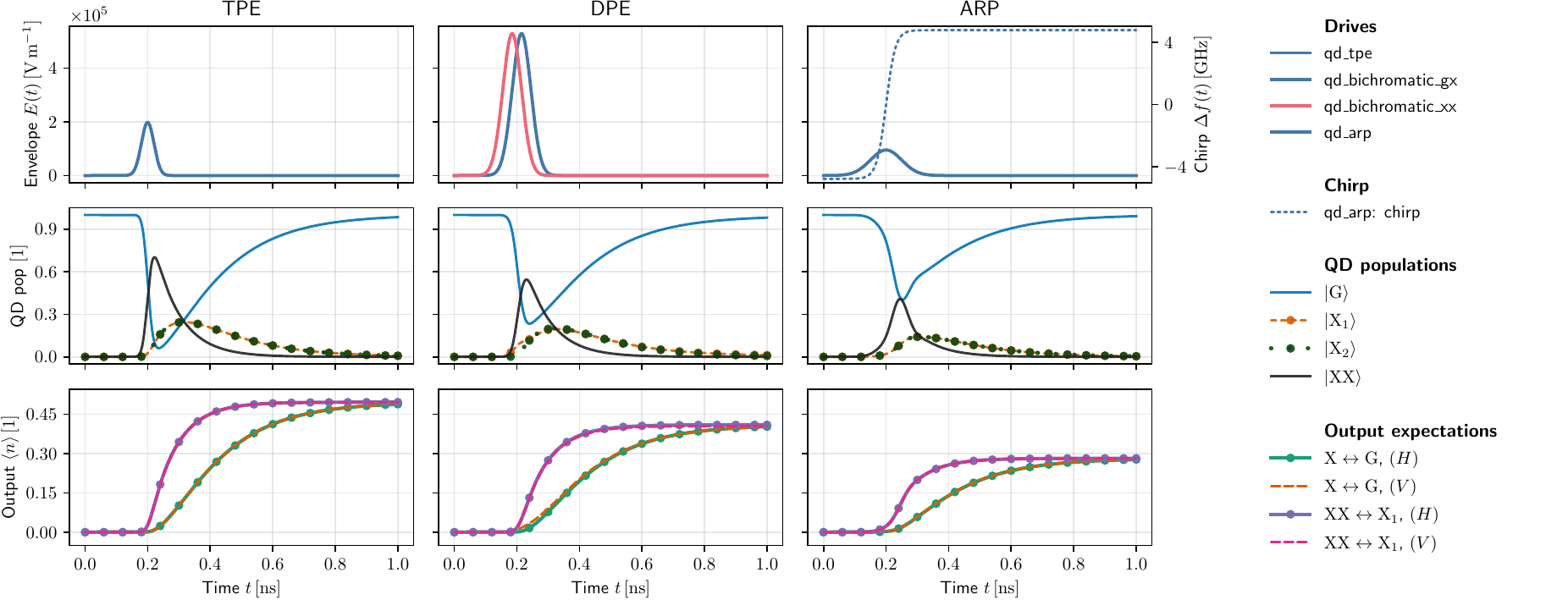}
    \caption{Different excitation schemes. From left to right: Resonant \gls{TPE} $\pi$-pulse, \gls{DPE} with two pulses shifted by $\pm \Delta_{\text{DPE}} = \pm \qty{31.83}{\giga\hertz}$ from the transition, \gls{ARP} chirped excitation with chirp function $\Delta f(t) = \qty{3.0e10}{} \tanh\left( (t - \qty{0.2}{\nano\second})/\qty{0.03}{\nano\second} \right)$. Phonon dissipation mechanisms were turned off for these simulations.}
    \label{fig:plot_schemes}
\end{figure}

Here we present some of the simulation outcomes with the resulting metrics to illustrate the capabilities of the simulation.
The \gls{QD} for both sets of plots (Figure \ref{fig:plot_tpe} and Figure \ref{fig:plot_schemes}) was configured with the following energy structure (\texttt{EnergyStructure}): $E_{\mathrm{X}} = \qty{1.3}{\electronvolt}$, $E_{\text{b}} = \qty{3}{\milli\electronvolt}$, $\Delta=\qty{5}{\micro\electronvolt}$.

\twocolumn

\begin{table*}
  \centering
  \caption{Summary of metrics corresponding to simulations shown in Figure \ref{fig:plot_tpe} and Figure \ref{fig:plot_schemes}. 
  Brightness is split into early/late photon numbers $(N_\text{early}, N_\text{late})$.
  Log-negativity $E_\mathcal{N}$ is reported both unconditional and conditional (post-selected).}
  \label{tab:metrics_grouped}
  \vspace{0.5em}

  \begin{tabular}{@{}lcccccccc@{}}
    \toprule
    \multicolumn{1}{c}{} & \multicolumn{2}{c}{\textbf{Brightness}} & \multicolumn{2}{c}{\textbf{Log-negativity}} & \textbf{Purity} & \textbf{Indist.} & \textbf{Coherence} & \textbf{Phase}\\
    \cmidrule(lr){2-3} \cmidrule(lr){4-5}
    \textbf{Scenario} & \(N_\text{early}\) & \(N_\text{late}\) & \(E_\mathcal{N}\) & \(E_\mathcal{N}^\text{cond}\) & \(P\) & $\Lambda$ & $|\rho_{\pm,\mp}|$ & $\expval{\phi}\,[\si{rad}]$\\
    \midrule

$\pi$-pulse     &\qty{0.957}{}&\qty{0.940}{}&\qty{0.650}{}&\qty{0.707}{}&\qty{0.572}{}&\qty{0.626}{}&\qty{0.3160}{}&\qty{0.9}{}\\
$5\pi$-pulse     &\qty{1.148}{}&\qty{1.124}{}&\qty{0.566}{}&\qty{0.677}{}&\qty{0.462}{}&\qty{0.626}{}&\qty{0.2996}{}&\qty{1.1}{}\\
det.\ $\pi$-pulse     &\qty{0.149}{}&\qty{0.147}{}&\qty{0.121}{}&\qty{0.700}{}&\qty{0.744}{}&\qty{0.626}{}&\qty{0.3122}{}&\qty{0.9}{}\\
    \addlinespace

TPE $\pi$-pulse     &\qty{0.991}{}&\qty{0.976}{}&\qty{0.689}{}&\qty{0.708}{}&\qty{0.655}{}&\qty{0.626}{}&\qty{0.3165}{}&\qty{0.9}{}\\
DPE     &\qty{0.814}{}&\qty{0.803}{}&\qty{0.027}{}&\qty{0.035}{}&\qty{0.326}{}&\qty{0.626}{}&\qty{0.0123}{}&\qty{-1.3}{}\\
ARP     &\qty{0.563}{}&\qty{0.554}{}&\qty{0.427}{}&\qty{0.707}{}&\qty{0.403}{}&\qty{0.626}{}&\qty{0.3161}{}&\qty{0.9}{}\\
    \bottomrule
  \end{tabular}
\end{table*}

\subsubsection{Two Photon Excitation}

The exciton dynamics for the different presented driving scenarios of \gls{TPE} are summarized in Figure \ref{fig:plot_tpe} and the corresponding metrics are collected in Table \ref{tab:metrics_grouped}. Across these simulations, phonon-induced dissipation mechanisms were disabled in order to show the isolated coherent excitation pathways and their intrinsic limitations. Indistinguishability is computed from the decay rates and since all simulations used the same \gls{QD} configuration, the indistinguishability remains unchanged for all cases.

For \gls{TPE}, the $\pi$-pulse equivalent exhibits efficient population transfer to the biexciton state. The excitation could be improved by shortening the length of the pulse as inherent decay is a parallel dynamic which is active also during the pulse. The $5\pi$-pulse predictably overdrives the dot and the resulting Rabi oscillations are visible in the population dynamics. The detuned $\pi$-pulse looses the coupling strength due to the detuning so the population transfer is not efficient.

This behaviour is directly reflected in the metrics Table~\ref{tab:metrics_grouped}. The stronger pulse produces the most brightness while the detuned pulse produces the least. However, the gain in the photon yield does not translate to the improved state quality. The unconditional log-negativity is the best for the $\pi$-pulse which has a clean excitation and decay pathway, whereas in the $5\pi$-pulse case the log-negativity is reduced despite the increased brightness, as the strong driving induces re-excitations and transient population of intermediate exciton states, leading to incoherent admixtures in the emitted state.

Additionally the re-excitation mechanism introduces additional dynamical phase as reflected in the increased average phase $\expval{\phi}$. While the constant phase offset does not in itself reduce entanglement, it is indicative of emission occurring over an extended time window during the pulse. This temporal spread also manifests a slight reduction in the polarization coherence $\lvert \rho_{\pm, \mp}\rvert$, consistent with the presence of multiple emission pathways and partial which-path information.

Overall, these results demonstrate that within resonant \gls{TPE}, increasing the pulse area primarily enhances the brightness at the cost of reduced purity and unconditional entanglement, while the intrinsic polarization entanglement of the cascade remains largely preserved.

\subsubsection{DPE and ARP Chirped Excitation Schemes}

Taken together, Figure \ref{fig:plot_schemes} and Table~\ref{tab:metrics_grouped} reveal clear and distinct trade-offs between the investigated excitation schemes. Resonant \gls{TPE} provides the most favourable balance between brightness and entanglement quality, combing high photon yield with large unconditional and conditional log-negativity.

In contrast \gls{DPE} shows a pronounced reduction in entanglement. This degradation can be traced back to the fact that decay largely occurs during the drive due to the used timescales for excitation and decay. This introduces which-path information in the sense that the excitation pulse acts also as a stimulation pulse for the emission. Naturally, that destroys the coherence required for polarization entanglement in the subsequent cascade.

\gls{ARP} chirped excitation exhibits robust population transfer of the biexciton state, as evidenced by the smooth and monotonic population dynamics. However, the reduced brightness compared to resonant \gls{TPE} reflects the longer interaction time and the incomplete suppression of residual exciton occupation. While this performance is inferior to \gls{TPE} under idealized, phonon-free conditions considered here, \gls{ARP} offers a fundamentally different advantage. Due to the frequency sweeping, population transfer does not rely on precise resonance conditions. As a result, \gls{ARP} is intrinsically more robust against detuning, pulse-shape imperfections, and fluctuations in the system parameters.

\subsubsection{Phonons}

To showcase the effects of the phonons on the $\mathrm{XX}$-$\mathrm{X}$-cascade, we simulate the cascade under resonant \gls{TPE} while explicitly inducing phonon-induced decoherence. The \gls{QD} is coupled to an effective acoustic phonon bath. To make the temperature dependence of phononic effects directly visible in the emitted photon entanglement, we supplement the polaron dressing with a custom temperature-dependent phenomenological exciton relaxation channel between the two exciton eigenstates. The relaxation rates are constructed from a super-ohmic deformation-potential spectral density $J(\omega)$ and Bose-Einstein phonon occupation, yielding upward and downward scattering rates that increase monotonically with temperature. Physically, this models incoherent phonon-assisted population exchange and phase randomization between exciton states.
\begin{figure}
    \centering
    \includegraphics[width=0.95\linewidth]{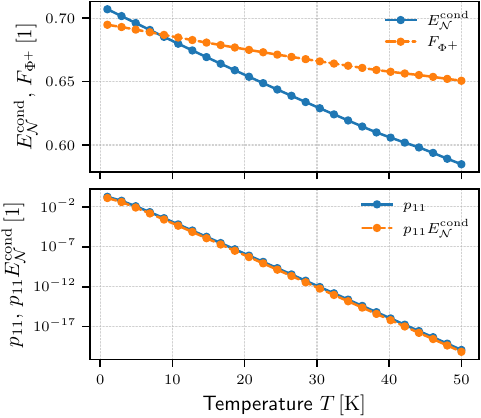}
    \caption{Temperature dependence of the two-photon entanglement yield under resonant \gls{TPE}. Shown is the product $p_{11}E_\mathcal{N}^\text{cond}$, where $p_{11}$ denotes the probability of detecting exactly one early and one late photon and $E_\mathcal{N}^\text{cond}$ is the conditional log-negativity of the post-selected polarization state. Increasing the structure's temperature leads to a monotonic reduction of the effective entanglement yield due to phononic effects.}
    \label{fig:plot-phonons}
\end{figure}

Figure \ref{fig:plot-phonons} illustrates the temperature dependence of the effective two-photon entanglement yield for resonant \gls{TPE}. The reducing entanglement yield is driven mainly by the polaron renormalization which decreases effective coupling. Reduced coupling leads to the imperfect population transfer into the biexciton state. This then manifests as a monotonic decrease of the two-photon emission probability $p_{11}$, and hence of the combined yield.

In addition to the reduced excitation efficiency, phonon-induced relaxation introduces a second, distinct degradation mechanism that affects the quality for the emitted entangled state itself. As summarized by Figure \ref{fig:plot-phonons}, the conditional log-negativity and Bell fidelity decrease with increasing temperature, even after post-selection on successful two-photon emission events.

Overall, phononic effects present in this simulation scenario impose a dual limitation on a resonant \gls{TPE}: a continuous reduction of excitation efficiency through polaron renormalization, and a temperature-activated degradation of polarization entanglement through exciton relaxation.

\subsubsection{Trends}

Across all investigated cases, the observed trends consistently link time-domain excitation dynamics to the resulting photon statistics and entanglement metrics. Excitation schemes that suppress intermediate excitation occupation and re-excitation during the drive simultaneously optimize brightness, purity, and unconditional log-negativity, while deviations from this ideal pathway introduce incoherent admixtures that predominantly reduce the usable fraction of emitted entangled photon pairs.

Beyond the comparison of individual excitation protocols, these results demonstrate the flexibility of the simulation framework to capture qualitatively different physical mechanisms within a unified model. Coherent excitation dynamics, multi-pulse interface, chirped frequency sweeps, and phonon-assisted incoherent processes are treated in the same manner.

The ability to simulate resonant, detuned, and chirped excitation schemes alongside temperature dependent phononic environments highlights the applicability of the framework to modelling realistic experimental conditions. Rather than optimizing a single excitation scenario, the presented results illustrate how the simulator enables systematic exploration of control strategies and error mechanisms.

\section{Conclusion}\label{section:Conclusion}

The parameters chosen for our simulations were selected to reflect realistic values reported in the literature. In particular, our assumptions are consistent with measurements presented in \cite{Scaparra2024, QuantumDotEntangledPhotonPairSource}, where comparable biexciton binding energies and \gls{FSS} have been observed. Thus, our simulation conforms reasonably to the expectation from the experiments. Additionally, the simulated dynamics reproduce the key features reported experimentally, supporting the validity of our approach. For a specific experiment users are encouraged to adapt specific values to match their material system, excitation scheme and experimental conditions. This will aid in achieving the best agreement between pre-experiment prediction and physical system behaviour.

We presented a unified simulation framework for modelling the excitation and emission dynamics of \gls{QD}-based $\mathrm{XX}$-$\mathrm{X}$-cascades under a wide range of optical control schemes. The framework enables direct simulation of coherent excitation protocols, including resonant, detuned, and chirped driving, as well as the incorporation of phonon-induced renormalization and incoherent relaxation processes, while providing access to both time-domain dynamics and emitted-photon observables.

Using this approach, we systematically compared resonant \gls{TPE}, \gls{DPE}, and \gls{ARP} chirped excitation, and quantified their impact on brightness, purity, and polarization entanglement. In the absence of phonons, resonant $\pi$-pulse excitation provides the most favourable balance between emission efficiency and entanglement quality, while stronger driving enhances brightness at the cost of increased incoherent admixtures. Yet, it has the most requirements with respects to precision of excitation pulse preparation. Chirped excitation schemes, although inferior in idealized conditions, exhibit increased robustness against detuning and parameter fluctuations, highlighting their relevance for realistic experimental settings.

By explicitly including phonon-induced effects, we further demonstrated that phonons impose a dual limitation on the entanglement generation process: a continuous reduction of excitation efficiency through polaron renormalization and a temperature-dependent degradation of intrinsic polarization coherence through exciton relaxation. Importantly, the latter mechanism cannot be mitigated by post-selection alone, underscoring the necessity of excitation schemes that are robust against both coherent control imperfections and environmental decoherence.

Overall, the presented results illustrate how the simulation framework enables systematic exploration of excitation strategies and error mechanisms in solid-state quantum light sources. This capability provides a foundation for optimizing entangled-photon generation under realistic operating conditions and for extending the model toward more complex photonic environments, detector models, and quantum-network-level simulations.

\section*{Acknowledgements}

We would like to thank Eduardo Zubizarreta Casalengua for very helpful general discussions and Davide Li Calsi for feedback and discussions on the mathematics section, and conceptualization of the introduction.

\subsection*{Data Availability}

The data that support the findings of this study are openly available on GitHub at \url{https://github.com/tqsd/BEC} under an open-source license, reference \cite{Dataset}.



\subsection*{Funding}

This work was supported in part by the DFG Emmy-Noether Program under Grant 1129/2-1, 
in part by the Federal Ministry of Research, Technology and Space under grant numbers 16KISQ039, 16KISR026, 16KIS1598K, 16KISQ093 and 16KISQ077, and in part by the Federal Ministry of Research, Technology and Space of Germany through the Programme of "Souverän. Digital. Vernetzt." Joint Project 6G-life, project identification number 16KISK002.
The generous support of the state of Bavaria via the 6GQT project is greatly appreciated. This research is part of the Munich Quantum Valley, which is supported by the Bavarian state government with funds from the Hightech Agenda Bayern Plus.

\subsection*{Contributions}

PK: Abstract, Sections \ref{section:Introduction}, \ref{section:Physics}, \ref{section:Conclusion},
improvements in and conceptualization of Sections \ref{section:Mathematics}, \ref{section:Simulation}. \\
SS: Sections \ref{section:Mathematics}, \ref{section:Simulation}, \ref{section:Conclusion}, Appendix \ref{appendix:rotated_ladder_operators}, simulation. \\
JN: Funding acquisition, project conceptualization and supervision.

\subsection*{Conflict of Interest Disclosure}

The authors declare no conflicts of interest.

\printbibliography

\newpage

\appendix{}
\section{Rotated Ladder Operators}\label{appendix:rotated_ladder_operators}

We consider a single spatio-temporal mode of light, which supports two orthogonal polarizations; horizontal ($\mathrm{H}$) and vertical ($\mathrm{V}$). The Hilbert space of such mode is the tensor product
\begin{equation}
    \mathcal{H} = \mathcal{F}_\mathrm{H}^{(d)} \otimes \mathcal{F}_\mathrm{V}^{(d)},
\end{equation}
where $\mathcal{F}_p^{(d)}$ is a truncated Fock space of dimension $d$ for polarization $p\in\{\mathrm{H,V}\}$. The truncation imposes a maximum photon number $n_\mathrm{max}=d-1$ in each polarization.

\subsection{Single-Polarization Ladder Operators}

Let $\hat a$ and $\hat{a}^\dagger$ denote the annihilation and creation operators on a single $d$-dimensional Fock space. In the number basis $\{\ket{n}\}_{n=0}^{d-1}$ they take the form:
\begin{align}
   \hat a &= \sum_{n=1}^{d-1}\sqrt{n}\op{n-1}{n}, & \hat{a}^\dagger&=\sum_{n=1}^{d-1}\sqrt{n+1}\op{n}{n-1}.
\end{align}

These satisfy the \textit{truncated} commutation relation:
\begin{equation}
[\hat{a},  \hat{a}^\dagger] = \hat{I}_d - d\op{d-1}{d-1},
\end{equation}
which differs from the infinite-dimensional case by the projector onto the top Fock level.

\subsection{Polarization-Resolved Operators}

We define the annihilation operators $\hat{a}_\mathrm{H}$ and $\hat{a}_\mathrm{V}$ acting on $\mathcal{H}$ by:
\begin{align}
    \hat{a}_\mathrm{H} &= \hat{a} \otimes \hat{I}_d, & \hat{a}_\mathrm{V}&=\hat{I}_d\otimes\hat{a}.
\end{align}

Their adjoints $\hat{a}_\mathrm{H}^\dagger$, $\hat{a}_\mathrm{V}^\dagger$ are defined analogously. From the single-mode commutator above it follows that
\begin{align}
\begin{split}\label{eq:commutators}
    [\hat{a}_\mathrm{H}, \hat{a}_\mathrm{H}^\dagger] &= \hat{I}_{d^2}-d(\hat{P}\otimes\hat{I}_d),\\
    [\hat{a}_\mathrm{V}, \hat{a}_\mathrm{V}^\dagger] &= \hat{I}_{d^2}-d(\hat{I}_d\otimes \hat{P}),\\
\end{split}
\end{align}
where $\hat{P}=\ket{d-1}\bra{d-1}$ is the top-level  projector in $\mathcal{F}^{(d)}$.

Furthermore,
\begin{equation}
    [\hat{a}_\mathrm{H}, \hat{a}_\mathrm{V}^\dagger] = [\hat{a}_\mathrm{V}, \hat{a}_\mathrm{H}^\dagger] = 0,
\end{equation}
since they act on different tensor factors.

\subsection{Rotated Polarization Basis}

A general orthogonal (unitary) transformation in the polarization subspace is given by an $\mathrm{SU}(2)$ rotation:
\begin{equation}
    \begin{pmatrix}
        \hat{a}_+\\
        \hat{a}_-
    \end{pmatrix}
    = \begin{pmatrix}
        \cos\theta & e^{+i\phi}\sin\theta\\
        -e^{-i\phi}\sin\theta & \cos\theta
    \end{pmatrix}
    \begin{pmatrix}
        \hat{a}_\mathrm{H}\\
        \hat{a}_\mathrm{V}
    \end{pmatrix},
\end{equation}
with $\theta\in[0,\pi / 2]$ and $\phi\in [0,2\pi)$.
The operators $\hat{a}_+$ and $\hat{a}_-$ correspond to annihilation in the rotated polarization modes ("plus" and "minus").

The corresponding creation operators are:
\begin{align}
    \hat{a}_+^\dagger(\theta)&=\cos(\theta) \hat{a}_\mathrm{H}^\dagger + e^{-i\phi}\sin(\theta) \hat{a}_\mathrm{V}^\dagger,\\
    \hat{a}_-^\dagger(\theta)&=e^{+i\phi}\sin(\theta) \hat{a}_\mathrm{H}^\dagger + \cos(\theta)\hat{a}_\mathrm{V}^\dagger.
\end{align}

\subsection{Commutation Relations in the Rotated Basis}

Using bilinearity and the fact that cross-polarization commutators vanish, we find:
\begin{align}
    [\hat{a}_+, \hat{a}_+^\dagger] &=\cos^2(\theta)[\hat{a}_\mathrm{H},\hat{a}_\mathrm{H}^\dagger]+\sin^2(\theta)[\hat{a}_\mathrm{V},\hat{a}_\mathrm{V}^\dagger],\\
    [\hat{a}_-, \hat{a}_-^\dagger] &=\sin^2(\theta)[\hat{a}_\mathrm{H},\hat{a}_\mathrm{H}^\dagger]+\cos^2(\theta)[\hat{a}_\mathrm{V},\hat{a}_\mathrm{V}^\dagger].
\end{align}

Explicitly, inserting the truncated commutators from Equations \eqref{eq:commutators}:
\begin{align}
\begin{split}
    [\hat{a}_+, \hat{a}_+^\dagger]  &= \hat{I}_{d^2}-d \Big(\cos^2(\theta) \big(\hat{P}\otimes\hat{I}_d\big) \\
    &+ \sin^2(\theta) \big(\hat{I}_d\otimes \hat{P}\big) \Big),
\end{split}
\\
\begin{split}
    [\hat{a}_-, \hat{a}_-^\dagger]  &= \hat{I}_{d^2}-d \Big(\sin^2(\theta) \big(\hat{P}\otimes\hat{I}_d \big) \\
    &+ \cos^2(\theta) \big(\hat{I}_d\otimes \hat{P}\big) \Big).
\end{split}
\end{align}

In the infinite-dimensional limit $d\to\infty$, the projector terms vanish and the canonical bosonic commutation relations $[\hat{a}_\pm,\hat{a}_\pm^\dagger]=\hat{I}$ are recovered.

\subsection{Choice of Polarization Rotation Angle for Quantum Dot Transitions}

In the context of semiconductor quantum dots, the two bright exciton states $\ketXone$ and $\ketXtwo$ correspond to dipole transitions emitting photons of orthogonal linear polarizations, which we identify with the $\mathrm{H}$ and $\mathrm{V}$ basis states of the photonic mode.

In the absence of \gls{FSS} ($\Delta=0$), the exciton manifold is degenerate, and any orthogonal polarization basis is equally valid. In this case the rotation parameters $(\theta, \phi)$ are arbitrary, and one often chooses $\theta=0$ or $\theta=\pi/2$ to align the photonic polarization basis directly with the exciton eigenstates.

When \gls{FSS} is non-zero ($\Delta\neq0$), the Hamiltonian of the two bright excitons (up to an overall shift $E_0$) can be written in the $\{\mathrm{H},\mathrm{V}\}$ basis as
\begin{equation}
    \hat{H}_{\mathrm{X}}=\begin{pmatrix}
        E_0+\frac{\Delta}{2} & \delta \\
        \delta^* & E_0-\frac{\Delta}{2}
    \end{pmatrix},
\end{equation}
where $\Delta \in \mathbb{R}$ and $\delta \in \mathbb{C}$ captures anisotropic mixing. Diagonalizing gives eigenenergies $E_\pm=E_{0}\pm \frac{\Omega}{2}$ with
\begin{equation}
\Omega=\sqrt{\Delta^2+4\lvert\delta\rvert^2}.
\end{equation}

The corresponding eigenstates are obtained by a rotation of the $\{\mathrm{H},\mathrm{V}\}$ basis by a real mixing angle $\theta$ and a phase $\phi=\arg(\delta)$. The angle $\theta\in[0,\pi/2]$ is fixed by
\begin{align}
    \tan(2\theta)&=\frac{2\lvert\delta\rvert}{\Delta},\\
    \cos\theta&=\sqrt{\frac{1}{2}\left(1+\frac{\Delta}{\Omega}\right)},\\
    \sin\theta&=\sqrt{\frac{1}{2}\left(1-\frac{\Delta}{\Omega}\right)}.
\end{align}

Identifying the rotated photonic modes ($+$,$-$) with these eigenvectors, the rotation parameters that define $\hat{a}_\pm$ in the previous subsection are thus:
\begin{align}
    \theta&=\frac{1}{2}\arctan\left(\frac{2\lvert\delta\rvert}{\Delta}\right),\\
    \phi&=\arg(\delta),
\end{align}
so that the "$+$" mode couples to the higher-energy exciton and "$-$" mode to the lower-energy exciton. Special cases follow immediately: if $\delta=0$ then $\theta=0$ and the lab basis is already the eigenbasis; if $\Delta=0$ then $\theta=\pi/4$ (maximal mixing) and $\phi$ sets the azimuth of the eigenmodes.

\end{document}